# Pressure dependent *ab initio* study of the physical properties of hexagonal BeB$_2$C: a possible high-$T_c$ superconductor

Ruman Ali, Md. Enamul Haque, Jahid Hassan, M. A. Masum, R. S. Islam and S. H. Naqib*
Department of physics, University of Rajshahi, Rajshahi 6205, Bangladesh
*Corresponding author email: salehnaqib@yahoo.com

**Abstract**

This study uses *ab initio* Density Functional Theory (DFT) to explore the pressure dependent properties of hexagonal BeB$_2$C. The pressure dependent comprehensive investigation meticulously examined the structural, mechanical, electronic, optical, thermophysical, and superconducting properties of BeB$_2$C, unveiling several significant outcomes. With increasing pressure, BeB$_2$C undergoes a contraction in lattice parameters and unit cell volume, while its mechanical properties exhibit pronounced anisotropy. Using the calculated elastic constants and phonon dispersion, the mechanical stability of the compound was predicted in accordance with the Born-Huang stability criteria for hexagonal structures. The Cauchy pressure and the Pugh's ratio values establish a pressure dependent brittle-ductile-brittle transition, offering insights into the dual nature of the material. The metallic nature of BeB$_2$C was substantiated at ambient pressure, with pressure induced alterations in electronic band structure and Fermi surface topology suggesting a potential for tunability across various applications. The phonon dispersion and phonon density of states show the dynamical stability under pressure. The thermophysical properties (Debye temperature, melting temperature, mean sound velocities, thermal expansion coefficient and heat capacity) are also investigated under varying pressure conditions. Finally, the exploration of superconducting properties found that the transition temperature ($T_c$) is in good agreement with previously reported values, and illustrated that BeB$_2$C holds considerable promise as a high-temperature superconductor, with pressure augmenting its superconducting properties.

## 1. Introduction

Metal borides provide the only systems, apart from the boron hydrides and their derivatives, in which extensive clustering and catenation of boron atoms occur. The bonding in such systems has presented some interesting problems, and in some instances, may be tentatively explained by analogy with the hydride systems. Borides are also arousing considerable technological interest, since they are refractory and often chemically inert [1-3]. Boron carbide, renowned for its formidable hardness, exceptional fracture toughness, and commendable oxidation resistance at a low density, is pivotal in the development of avant-garde multifunctional ceramic matrices and finds application as a reinforcement agent in composite armor systems, as well as in corrosion-resistant cladding for fusion reactors and high-temperature, radiation-resilient semiconductor components [4]. As one of the new possible ternary borides BeB$_2$C was investigated in the past few years as a possible high-temperature superconductor, and showed a good agreement with conventional superconductors (like MgB$_2$) whose mechanism is based on electron-phonon coupling. The boron atoms often form layers or networks that provide structural stability, while beryllium and carbon atoms influence the charge distribution and the electron pairing mechanisms responsible for superconductivity. The interplay between these atoms leads to specific effects, such as modulation of the electron-phonon interactions and enhancement of the superconducting critical temperature. The arrangement of Be atoms, in particular, tends to localize electron density, which can affect the superconducting gap and overall performance. $BeB_2C$ is gaining attention for high-tech applications that require efficient, high-performance materials capable of operating at relatively high temperatures with minimal energy loss [5-7]. From the investigated system it is observed that, $BeB_2C$ is trigonal crystal structure belonging to the hexagonal family. Where $\alpha$ is the angle between edges $b$ and $c$, $\beta$ is the angle between edges $a$ and $c$, and $\gamma$ is the angle between edges $a$ and $b$. The conditions that the lattice parameters of trigonal crystal system of hexagonal axes are $a = b \neq c$ and, $\alpha = \beta = 90º$, $\gamma = 120º$. The paper utilizes first-principles calculations based on density functional theory (DFT) to comprehensively investigate the pressure dependent properties of hexagonal $BeB_2C$. The workflow begins by optimizing the structural properties of $BeB_2C$, including lattice parameters, equilibrium volume, and final energy, under varying hydrostatic pressures (0-5 GPa). Following this, the mechanical properties, such as bulk modulus, shear modulus, Young's modulus, Poisson's ratio, and the elastic constants are determined across the same pressure range. Phonon dispersion curves are then calculated to explore vibrational properties, while the electronic band structure and Fermi surface topology are examined alongside total and partial density of states (DOS). The thermophysical properties are assessed through Debye temperature, wave velocities, anisotropy indices, melting temperature, heat capacity and thermal conductivity, all under varying pressures. Additionally, the Kleinman parameter, Grüneisen parameter, and thermal expansion coefficient are explored. Optical properties such as absorption coefficient, conductivity, dielectric function, refractive index, and reflectivity are investigated for electric field polarizations along [100] and [001] directions under pressure. Finally, superconducting properties, including the transition temperature ($T_c$), are evaluated using the electronic band structure and density of states, providing critical insights into potential high-temperature superconductivity under pressure. This systematic approach integrates structural, mechanical, electronic, thermophysical, optical, and superconducting analyses to explore the comprehensive behavior of $BeB_2C$ under pressure We expect that this study will enrich the comprehension of possible atmospheric superconductor $BeB_2C$ as a viable material for applications in superconductivity, optoelectronics, and advanced materials, establishing a foundational basis for subsequent experimental and theoretical explorations on ternary metallic-borides. The remaining parts of this paper are organized as follows. In Section 2, the computational methodology using the DFT-based approach with CASTEP is discussed, including details on the computational parameters and methods used to calculate the elastic, optical, and thermophysical properties. The results and discussions are presented in Section 3, where the structural, elastic, electronic, optical, thermophysical, and superconducting properties of $BeB_2C$ under varying pressures are systematically analyzed. Finally, Section 4 provides a summary of the key findings and conclusions of the present work.

## 2. Computational methodology

In the present study, the stable structure of $BeB_2C$ was obtained using the method of structural optimization, and the detail optimization method was implemented by the density functional theory (DFT) with the Local Density Approximation (LDA) in the CAmbridge Serial Total Energy Package (CASTEP) [8-10]. The interactions between ionic core and valence electrons are characterized by the (OTFG) ultra-soft pseudopotential generalized by Perdew-Burke-Ernzerhof (PBE) scheme.

The cutoff energy of the plane wave [11] is selected as 500 eV and the $\Gamma$-center Monkhorst-Pack $k$-points grid of 27×27×7 was chosen. The Monkhorst-Pack special point's scheme for the Brillouin zone (BZ) integration was determined, according to our experience, based on the size of crystal lattice structure [12].

Figure 1(a) and 1(b) show the hexagonal BeB$_2$C atomic crystal structure. In this cell, Be atom is located at the site (0.00000 0.00000 −0.25997), two B atoms occupy the sites (0.00000 0.00000 −0.68127), (0.00000 0.00000 −0.42377), and C is located at the site (0.00000 0.00000 −0.55254) [5]. Here, we exclusively considered the valence electrons, corresponding to the electronic configurations: Be [$2s^2$], B [$2s^2\ 2p^1$], and C [$2s^2\ 2p^2$]. By focusing solely on the valence states, which play a critical role in bonding and material properties, we ensured that the calculations captured the essential interactions governing the structural and electronic behavior of these elements. This approach not only simplifies the computational model but also enhances its accuracy in predicting material characteristics under varying conditions [13]. The Broyden-Fletcher-Goldfarb-Shanno (BFGS) minimization technique [11] has been adopted to optimize the crystal structure of space group $R$3m (No. 160) with hexagonal axes and convergence tolerance is set to ultra-fine quality with energy change below $5.0\times10^{-6}$ eV/atom, force less than 0.01 eV/Å, stress less than 0.02 GPa, and change in atomic displacement less than $5.0\times10^{-4}$ Å. The quadratic elastic constants and the corresponding elastic properties have not been examined in any other prior research. The elastic constants were calculated by the 'stress-strain' method contained in the CASTEP. Voigt-Reuss-Hill (VRH) averaging scheme [14] was used for calculating the polycrystalline mechanical parameters from the estimated $C_{ij}$. The polycrystalline elastic moduli, Pugh's ratio, Poisson's ratio, universal elastic anisotropy, and Vickers hardness were evaluated by using the corresponding equations given elsewhere. The prediction of whether BeB$_2$C will be brittle or ductile is conducted through Pugh or Pettifor criterion. To obtain the smooth Fermi surface, 45×45×10 $k$-point mesh has been used.

Using the phonon calculation by the perturbative linear response theory and the ultrasoft pseudopotentials, we investigate the contribution of vibration to the thermodynamics parameters [15,16]. The optical properties are determined implying 5.0 eV plasma energy and a Drude damping of 0.05 eV. The frequency-dependent dielectric function $\varepsilon(\omega)$ represents the interaction of incident photon and explains the dispersion and absorption of light energy. The real part $\varepsilon_1(\omega)$ and the imaginary part $\varepsilon_2(\omega)$ of the dielectric constant are calculated using Kramers-Kronig relationships [13] as follows:

$$\varepsilon_1(\omega) = 1 + \frac{2}{\pi}P\int_0^\infty \frac{\omega'\varepsilon_2(\omega')}{\omega'^2 - \omega^2}d\omega \tag{1}$$

$$\varepsilon_2(\omega) = \frac{2e^2\pi}{\Omega\varepsilon_0}\sum_{k,v,c}|\langle\psi_k^c|\hat{u}\times r|\psi_k^v\rangle|^2\delta(E_k^c - E_k^v - E) \tag{2}$$

where $\omega$ is the light frequency, $e$ is the electronic charge, and $\psi_k^c$ and $\psi_k^v$ are the conduction band (CB) and valence band (VB) wavefunctions at $k$, respectively. The investigated electronic band structure of BeB$_2$C is used in this formula. All other optical constants on the energy dependence of the absorption spectrum, the refractive index, the extinction coefficient, the energy-loss spectrum, and the reflectivity can be derived from $\varepsilon_1(\omega)$ and $\varepsilon_2(\omega)$ [17] using the CASTEP tool based on the standard DFT Kohn-Sham orbitals [18]. In this work, the superconducting transition temperature has been evaluated using McMillan equation [19] by utilizing relations between superconducting transition temperature, $T_c$, Debye temperature, $\theta_D$ and electron-phonon coupling constant, $\lambda$. The Debye temperature and other thermophysical parameters are estimated from the elastic properties and elastic moduli of BeB$_2$C.

For a better understanding of bonding characteristics, the Mulliken bond population analysis and charge density distribution have been utilized to study the bonding properties of solids. BeB$_2$C makes advantage of the projection of the plane-wave states onto a linear combination of atomic orbital (LCAO) basis sets [20]. To perform the Mulliken bond population analysis, the Mulliken density operator written on the atomic basis is utilized as:

$$P^M_{\mu\nu}(g) = \sum_{g'}\sum_{v'} P_{\mu v'}(g') S_{v'v}(g-g') = L^{-1} \sum_k e^{-ikg} (P_k S_k)_{\mu v'} \tag{3}$$

In this study, the third-order Birch-Murnaghan equation of state [21] was utilized to generate the energy-volume data based on zero temperature and zero pressure equilibrium values for energy, volume, and bulk modulus, which were derived from DFT calculations. Also, the bond hardness of the covalent crystal has been computed using well known semi-empirical formula. This approach allows for a detailed investigation into the thermodynamic, elastic, and structural properties that vary with changes in pressure. By employing the Birch-Murnaghan equation, the study captures the relationship between energy and volume under different pressure conditions, providing insights into how the properties of a material evolve under compression. This methodology is particularly effective in understanding the responses of the material to external pressures and their corresponding mechanical and structural behaviors.

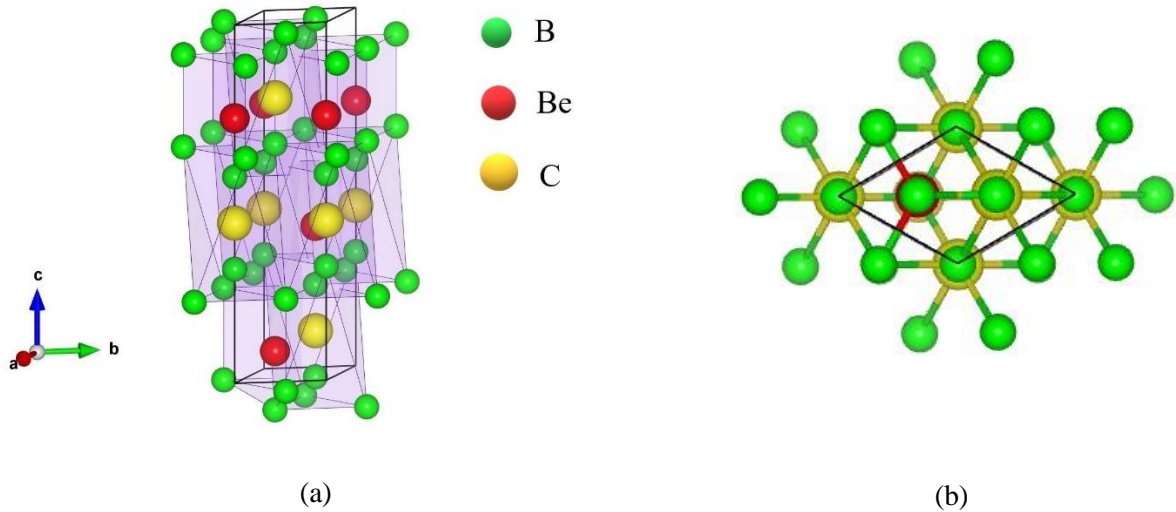

(a) (b)

**Figure 1** (a) Crystal structure of Hexagonal BeB$_2$C, (b) Top view.

## 3. Results and discussion

### 3.1 Structural properties

In order to achieve structural analysis, such as determining the lattice constant, $a$ (Å), bulk moduli, $B$ (GPa), and its pressure responses, the volume of the unit cell of $BeB_2C$ was tuned by optimization of the parameters. The geometry optimization of ternary $BeB_2C$ was performed using density functional theory (DFT) with the Ceperley and Alder data as parameterized by Perdew and Zunger (CA-PZ) in the framework of the local density approximation (LDA) as implemented in the Cambridge Serial Total Energy Package (CASTEP) plane-wave code. Table 1 contains the optimized lattice parameters as well as the theoretical lattice parameters that are currently available. The table shows that as pressure is steadily increased, both the cell volume and the values of the lattice parameters gradually decrease. From Table 1, it can be seen that the optimum lattice volume of $BeB_2C$ was 87.60 Å$^3$, which was compressed by 1.95% as the pressure was increased from 0 GPa to 5 GPa. This is to be expected since the crystal is compressed by the pressure rise, which lowers the cell volume and related lattice properties. Density mixing has been used to the electronic structure and the Broyden Fletcher Goldfarb Shanno (BFGS) geometry optimization has been adopted to optimize the trigonal structure of hexagonal family belonging to the space group $R\bar{3}m$ (No. 160) with hexagonal axes and the number of atoms per unit cell is 12 (B = 6, Be = 3 and C = 3).

The data from the Table 1 indicate that under increasing pressure, the $BeB_2C$ structure undergoes anisotropic compression, with the most significant changes occurring along the $c$-axis at 2 GPa, while the $a$-axis shows a steady but less pronounced reduction. The overall volume also consistently decreases, reflecting the response of material to external pressure.

**Table 1** Calculated lattice constants ($a = b$ in Å) and $c$, $c/a$ ratio and equilibrium cell volume ($V$ in Å$^3$) of $BeB_2C$ structure corresponding to applied pressure ($P$ in GPa).

| $P$ | $a = b$ | $c$ | $c/a$ | $V$ | Ref. |
|---|---|---|---|---|---|
| - | 2.96 | 11.51 | 3.88 | 87.60 | [5] |
| 0.0 | 2.97 | 11.51 | 3.88 | 88.11 | [This work] |
| 1.0 | 2.96 | 11.50 | 3.88 | 87.64 | |
| 2.0 | 2.96 | 11.41 | 3.86 | 87.19 | |
| 3.0 | 2.95 | 11.48 | 3.89 | 86.74 | |
| 4.0 | 2.95 | 11.47 | 3.89 | 86.31 | |
| 5.0 | 2.94 | 11.46 | 3.89 | 85.89 | |

### 3.2 Elastic properties

Investigations of elastic constants are vital for applications related to the mechanical properties of solids. They provide information on stability, bonding, ductility, brittleness, anisotropy, compressibility, Vicker's hardness, and stiffness of solids. There are six independent elastic tensors ($C_{11}$, $C_{33}$, $C_{44}$, $C_{12}$, $C_{13}$, and $C_{14}$) for the rhombohedral (I) class (Laue class $\bar{3}m$) [22] of $BeB_2C$ crystal structure. These constants provide great assistance in determining the material strength regarding the response of the crystal to external forces. For a material of hexagonal axes symmetry, the stiffness constant $C_{66}$ depends [23] on the two other independent stiffness constants $C_{11}$ and $C_{12}$ via the expression, $C_{66} = (C_{11} - C_{12})/2$. All the calculated elastic stiffness constants at different pressures are listed in Table 2 for BeB$_2$C. The Born-Huang criterion [18] can be used to analyze mechanical stability of BeB$_2$C structure. For hexagonal system, the criteria are:

$$C_{11} > |C_{12}|; \; C_{44} > 0$$

$$C_{13}^2 < \frac{1}{2}C_{33}(C_{11} + C_{12})$$

$$C_{14}^2 + C_{15}^2 < \frac{1}{2}C_{44}(C_{11} - C_{12}) = C_{44}C_{66}$$

(4)

**Table 2** The calculated single crystal elastic constants ($C_{ij}$ in GPa), Cauchy pressures (*CP* in GPa) and machinability index ($\mu_M$) under different pressures (*P* in GPa).

| P | $C_{11}$ | $C_{12}$ | $C_{13}$ | $C_{14}$ | $C_{33}$ | $C_{44}$ | $C_{66}$ | CP | $\mu_M$ | Ref. |
|---|---|---|---|---|---|---|---|---|---|---|
| 0.0 | 307.40 | 119.68 | 25.14 | 23.18 | 769.96 | 145.63 | 93.86 | −26.03 | 1.26 | |
| 1.0 | 312.62 | 124.19 | 29.85 | 25.32 | 779.55 | 146.98 | 94.22 | −22.79 | 1.28 | |
| 2.0 | 319.00 | 126.66 | 31.91 | 24.66 | 787.58 | 148.30 | 96.67 | −21.64 | 1.30 | [This work] |
| 3.0 | 296.14 | 148.25 | 56.01 | 24.45 | 736.18 | 123.47 | 73.95 | 24.78 | 1.60 | |
| 4.0 | 301.09 | 152.48 | 62.20 | 25.84 | 738.44 | 125.74 | 74.30 | 26.74 | 1.61 | |
| 5.0 | 315.89 | 122.44 | 28.14 | 23.75 | 778.54 | 147.23 | 96.72 | −24.79 | 1.28 | |

Table 2 lists the single crystal elastic constants $C_{ij}$ and Cauchy pressures [24] for BeB$_2$C under various external pressures ranging from 0 to 5 GPa. It is seen that the mechanical stability criteria are satisfied which confirms that the compound is mechanically stable as well. The elastic constants, which are fundamental to understanding the mechanical behavior of a material, provide insight into its response to stress and deformation in different crystallographic directions. In light of the absence of theoretical findings in existing literature, we anticipate that our projections may serve as a foundational reference for subsequent investigations pertaining to BeB$_2$C.

Elastic constants $C_{11}$, $C_{12}$, and $C_{13}$ are the principal longitudinal elastic constants, representing the stiffness along specific axes. As pressure increases, $C_{11}$ demonstrates a progressive rise, starting from 307.40 GPa at 0 GPa and reaching 315.89 GPa at 5 GPa. This increase signifies a strengthening the rigidity of the crystal along the principal axis, suggesting that BeB$_2$C becomes stiffer under higher pressure. The elastic constants $C_{12}$ and $C_{13}$ exhibit a similar trend, with $C_{13}$ showing the most significant increment, rising from 25.14 GPa at 0 GPa to 62.20 GPa at 4 GPa. This marked increment highlights the enhanced resistance of the material to deformation in the corresponding crystallographic planes under increasing external pressure. From Table 2 it is seen that, for all the pressures the compound BeB$_2$C shows $C_{33} > C_{11}$. This indicates that the crystal offers more resistance along *c*- direction than *a*- and *b*- direction according to uniaxial strain values. This mainly signifies the layered nature of BeB$_2$C under the investigation of pressure response. Also, the chemical bonding in the *c*- direction is stronger than *ab*-plane.

The stiffness constant $C_{44}$ is used to represent the resistance offered by the compound to the shear deformation with respect to a tangential stress applied across the (001) plane in the [010] direction. It is clearly seen that from Table 2, the $C_{11}$ and $C_{33}$ values are noticeably larger than those of $C_{12}$, $C_{13}$, $C_{44}$ and $C_{66}$, which characterize the resistance to shear deformations, indicating that the studied compound is more resistant to compression than to shear. Consistent with Table 2, the value of $C_{12}$ is higher than $C_{44}$ for the pressures 3 GPa and 4 GPa, which indicates that the compound is more resistant to shear than to extension [25]. Since, in all cases from Table 2 it can be seen that, this condition is satisfied and this indicates that the compound which is being studied is more susceptible to shear deformations. However, $C_{12}$, $C_{13}$, and $C_{44}$ are related to the shape elasticity.

The Cauchy pressure ($CP = C_{12} - C_{44}$), a parameter indicative of the nature of bonding in the material, varies significantly with applied pressure. At ambient conditions (0 GPa), the Cauchy pressure is negative (−26.03 GPa), characteristic of predominantly covalent bonding in the crystal [26]. As pressure increases, the $CP$ gradually becomes less negative, and at 4.0 GPa, it even turns positive (26.74 GPa). This shift toward positive values suggests a gradual transition in bonding nature, with the material exhibiting increased metallic characteristics at higher pressures. The positive Cauchy pressure at elevated pressures indicates a significant alteration in the electronic environment, which could have implications for the conductivity and ductility of the material.

Machinability index ($\mu_M = B/C_{44}$) is a measure of the ease with which a material can be machined. At 0 GPa, BeB$_2$C has a modest machinability of 1.26, which improves significantly at higher pressures. The machinability peaks at 4 GPa with a value of 1.61, suggesting that BeB$_2$C becomes easier to machine as pressure increases up to this point. At 5 GPa, however, $\mu_M$ drops back to 1.28, indicating reduced machinability at extreme pressures.

The shear modulus $C_{66} = (C_{11} - C_{12})/2$, which quantifies the resistance of material to shear deformation [27], shows an increase with applied pressure. At 3 GPa, $C_{66}$ reaches 73.95 GPa, increasing to 74.30 GPa at 4 GPa. This trend reflects an enhancement in the resistance of the material to shear stress, further supporting the conclusion that BeB$_2$C becomes mechanically more robust under higher pressures. Moreover, the slight increase in $C_{66}$ between 3.0 and 4.0 GPa suggests that the crystal experiences incremental improvements in its ability to withstand shear forces.

In terms of structural phase transitions, the data do not provide clear evidence of a phase transition, which would typically manifest as abrupt changes in the elastic constants. Instead, the gradual and continuous increases in $C_{11}$, $C_{12}$, and $C_{13}$ suggest that the material retains its crystalline structure across the examined pressure range. However, the transition from negative to positive Cauchy pressures points to a subtle shift in the bonding nature, from covalent to more metallic, particularly above 4 GPa. This suggests that while no significant structural phase transition occurs, a second-order transition may be taking place, where changes in electronic and mechanical properties evolve smoothly without a complete change in the crystal structure [28].

However, the elastic constants and Cauchy pressures of BeB$_2$C demonstrate that this material becomes increasingly stiffer and more resistant to shear deformation with rising external pressure. The transition from negative to positive Cauchy pressures further implies a shift in bonding nature, with metallic bonding becoming more prominent at higher pressures. These results are crucial for understanding the potential applications of BeB$_2$C in high-pressure environments, where enhanced mechanical stability and altered electronic properties may be desirable.

### 3.3 Mechanical properties

The elastic constants $C_{ij}$ reflect the degree to which crystals can return to their initial configuration subsequent to the removal of stress, provided that such stress remains within the elastic limit; thus, they constitute essential mechanical parameters of crystalline materials. Through the computation of these elastic constants, it becomes feasible to ascertain various mechanical properties, encompassing Young's moduli, Shear moduli, and Poisson's ratio. To estimate these properties, foremost two schemes of approximations are typically used, that are Voigt (V) and Reuss (R) [29,30]. According to Voigt approximation from the linear combination of various elastic constants, the isotropic bulk ($B_V$) and shear modulus ($G_V$) can be calculated based on the assumption of a homogeneous stress throughout the crystal. On the other hand, Reuss derives a different estimate for isotropic bulk ($B_R$) and shear modulus ($G_R$) from the single crystal elastic constants based on the assumption of a spatially homogeneous strain [31]. Later Hill proved that, the Voigt-Reuss-approximated values gives the lower limit of the polycrystalline elastic moduli whereas the real values lie between the Voigt and Reuss bounds [32]. The Hill averaging method defines the isotropic elastic moduli by considering the average of Voigt and Reuss parameters.

The Voigt and Reuss bounds of $B$ and $G$ for the hexagonal symmetry expressed as follows:

$$B_V = \frac{[2(C_{11} + C_{12}) + 4C_{13} + C_{33}]}{9} \tag{5}$$

$$\frac{1}{(B_R - C_{13})} = \frac{1}{(C_{11} - C_{66}) - C_{13}} + \frac{1}{(C_{33} - C_{13})} \tag{6}$$

$$G_V = \frac{1}{5}\left[\left(\frac{C_{11} + C_{33} - 2C_{13} - C_{66}}{3}\right) + 2C_{44} + 2C_{66}\right] \tag{7}$$

$$G_R = \left[\frac{1}{5}\left\{\frac{9B_V}{B_R(C_{11} + C_{33} - 2C_{13} - C_{66})^2} + \frac{2}{C_{44}} + \frac{2}{C_{66}}\right\}\right]^{-1} \tag{8}$$

Hill took an arithmetic mean:

$$B = \frac{1}{2}(B_V + B_R) \tag{9}$$

$$B = \frac{1}{2}(B_V + B_R) \tag{10}$$

By using following relations Young's modulus ($Y$), and Poisson's ratio ($n$) can be calculated as:

$$Y = \frac{9BG}{(3B + G)} \tag{11}$$

$$n = \frac{(3B - 2G)}{(6B + 2G)} \tag{12}$$

The computed Bulk modulus, Young's modulus, Shear modulus, Pugh's ratio and Poisson's ratio [33] of BeB$_2$C as defined in Equations (5)-(12) are listed in Table 3 for pressure range from 0 GPa to 5 GPa. This research primarily concentrates on examining the influence of pressure on various characteristics of the compound.

**Table 3** Elastic moduli ($B$, $G$ & $Y$; all in GPa), Pugh's ratio ($B/G$), Poisson's ratio ($n$), Vickers hardness ($H_V$ in GPa), and fracture toughness ($K_{IC}$ in MPam$^{0.5}$) of BeB$_2$C along with the values due to pressures ($P$ in GPa).

| P | $B_V$ | $B_R$ | B | $G_V$ | $G_R$ | G | Y | B/G | n | $H_V$ | $K_{IC}$ | Ref. |
|---|---|---|---|---|---|---|---|---|---|---|---|---|
| 0 | 191.63 | 175.50 | 183.56 | 158.01 | 125.04 | 141.53 | 337.78 | 1.29 | 0.19 | 29.33 | 34.00 | [This work] |
| 1 | 196.95 | 180.51 | 188.73 | 159.03 | 124.95 | 141.99 | 340.57 | 1.32 | 0.20 | 28.38 | 34.50 | |
| 2 | 200.94 | 184.64 | 192.79 | 161.12 | 127.75 | 144.44 | 346.73 | 1.33 | 0.20 | 28.89 | 35.14 | |
| 3 | 205.44 | 189.56 | 197.50 | 135.39 | 99.77 | 117.58 | 294.33 | 1.68 | 0.25 | 19.62 | 32.06 | |
| 4 | 210.48 | 194.75 | 202.53 | 136.07 | 100.04 | 118.05 | 296.55 | 1.71 | 0.25 | 19.77 | 32.50 | |
| 5 | 196.42 | 180.40 | 188.41 | 160.34 | 127.73 | 144.03 | 344.33 | 1.30 | 0.19 | 29.90 | 34.59 | |

Table 3 presents the bulk modulus ($B$), shear modulus ($G$), and Young's modulus ($Y$) of BeB$_2$C, as well as Pugh's ratio (B/G), Poisson's ratio ($n$), and other related mechanical properties under various pressure from 0 to 5 GPa. These moduli and ratios are essential in evaluating the ductility, brittleness, and overall mechanical stability of the material.

The bulk modulus is a constant within the elastic limit that expresses the resistance of a substance to uniform compression. The table shows that the bulk modulus increases from 183.56 GPa at 0 GPa to 202.53 GPa at 4 GPa, indicating that BeB$_2$C becomes more resistant to volumetric compression under increasing pressure. This trend is observed both for the Voigt approximation ($B_V$) and Reuss approximation ($B_R$), with $B_V$ and $B_R$ reaching their maximum values at 4 GPa before slightly decreasing at 5 GPa. The increase in bulk modulus under pressure is consistent with a general stiffening of the material, which is expected as the atoms in the crystal are pushed closer together, strengthening their bonds. The peak at 4 GPa indicates that BeB$_2$C is at its most incompressible state at this pressure, after which its resistance decreases, potentially signaling the onset of structural adjustments or bond weakening.

The Shear modulus is another parameter to characterize the strength of the material. It also represents the powerful predictor of hardness. The shear modulus ($G$) at 3 GPa reaches 117.58 GPa, reflecting a significant reduction compared to previous pressures, followed by a slight increase to 118.05 GPa at 4 GPa. However, at 5 GPa, the shear modulus jumps sharply to 144.03 GPa. This behavior suggests that, while the material softens slightly under shear stress at 3-4 GPa, there is a strong recovery in shear resistance at 5 GPa, possibly due to the reorientation of the crystal structure or a subtle phase adjustment that strengthens its shear response.

Young's modulus follows a similar trend, decreasing to 294.33 GPa at 3 GPa and slightly recovering to 296.55 GPa at 4 GPa. At 5 GPa, $Y$ rises sharply to 344.33 GPa, mirroring the behavior of the shear modulus. This recovery suggests an overall stiffening of the material in response to tensile stress at 5 GPa, implying that while the material softens slightly at intermediate pressures, it regains stiffness beyond this point.

The Pugh's ratio, which is the ratio between the bulk modulus, $B$ and shear modulus, $G$ (i.e. $B/G$): a qualitative indicator to categorize the brittleness and ductility behaviors of materials. According to Pugh, $B/G < 1.75$ indicates that the compound is considered brittle. Otherwise, it is ductile ($B/G > 1.75$) [34]. Pugh's ratio ($B/G$) is a critical parameter for understanding ductility versus brittleness. At 3 GPa, $B/G$ reaches 1.68, and at 4 GPa, it peaks to 1.71, indicating that BeB$_2$C becomes more ductile under these pressures. This ratio suggests ductile behavior. However, at 5 GPa, the $B/G$ ratio drops drastically to 1.30, suggesting that the material becomes brittle again at higher pressures. This behavior suggests that while BeB$_2$C becomes more ductile under moderate pressures (3-4 GPa), it transitions back to a brittle state at higher pressures (5 GPa), likely due to structural rearrangements that decrease its deformability.

In addition, Frantsevich et al. [35] considered Poisson's ratio ($n$) to distinguish between brittleness and ductility in a material, where critical value is 0.26. The compound is brittle when $n$ is lower than the value but if the value of $n$ is larger than 0.26 then the material is ductile. From the predicted $n$ values, it can be seen that BeB$_2$C is brittle in nature under pressure response. Poisson's ratio is a critical parameter in the evaluation of numerous mechanical characteristics of crystalline materials. It serves as an indicator of the resilience of solids in response to shear forces. A low value of $n$ signifies a robust stability against shear deformation. Furthermore, Poisson's ratio is intricately linked to the characteristics of interatomic interactions within solids. In materials where central force interactions predominate, $n$ typically falls within the range of 0.25 to 0.50, while for systems governed by non-central forces, $n$ is observed to extend beyond this range [35-37]. From the Table 3, the Poisson's ratio increases from 0.19 at 0 GPa to 0.25 at 3 and 4 GPa, indicating increased ductility and less covalent bonding. At 5 GPa, it falls back to 0.19, coinciding with the reduction in $B/G$, supporting the notion that the material transitions back to a more brittle state at higher pressures.

Vickers hardness ($H_V$), and fracture toughness ($K_{IC}$) are also important mechanical parameters. These values provide insights into the performance of the material in various applications. Following equations were used to calculate $H_V$ and $K_{IC}$ [38,39]:

$$H_V = \frac{Y(1-2n)}{6(1+n)} \quad (13)$$

$$K_{IC} = V_0^{1/6} G \left(\frac{B}{G}\right)^{1/2} \quad (14)$$

Vickers hardness is a measure of the resistance of the material to plastic deformation. The hardness increases steadily from 29.33 GPa at 0 GPa to a peak of 29.90 GPa at 5 GPa, indicating greater resistance to indentation and wear with increasing pressure. However, the hardness decreases sharply to 19.62 GPa at 3 GPa and 19.77 GPa at 4 GPa, suggesting a softening effect and at this point the material becomes highly machinable. Hardness of BeB$_2$C is quite high but it is not superhard. Solids with Vickers hardness $H_V > 40$ GPa are graded as superhard.

Fracture toughness, which assesses the resistance of material to crack propagation [40], remains fairly constant across the pressure range considered, with slight variation from 34.00 MPam$^{0.5}$ at 0 GPa to 34.59 MPam$^{0.5}$ at 5 GPa. This consistency implies that while BeB$_2$C becomes harder under moderate pressures, its ability to resist fracturing remains stable.

The analysis of the mechanical properties of BeB$_2$C under varying pressures, as presented in Table 3, reveals a significant pressure induced mechanical transition in the behavior of the material from brittle to ductile within a specific pressure range. This brittle-ductile-brittle transition under varying pressures provides critical insights into the mechanical stability of BeB$_2$C in different high-pressure environments.

The bulk moduli for the single crystal along different crystallographic axes have been evaluated and are shown in Table 4. $B_{\text{relax}}$ is the single crystal isotropic bulk modulus, which has the same value we found from the Reuss approximation. $\alpha$ and $\beta$ can be characterized as the relative change of $b$ and $c$ axis as a function of the deformation of the $a$ axis, while for hexagonal crystal $\alpha = 1$. The linear bulk modulus along the crystallographic axes can also be obtained from the pressure gradient. The values of $B_a$ and $B_b$ are same and small compared with the value of $B_c$, which suggests that the compound is much stiffer in the $c$ direction and it is more compressible along $ab$-plane. From the principles of elastic theory, by examining the characterization of the bulk modulus under the condition that the strains orthogonal to the axes of applied stress are uniform, particularly in the context of the response to hydrostatic pressure, one can deduce that [41]:

$$B_a = a\frac{dP}{da} = \frac{\Lambda}{1 + \alpha + \beta} \tag{15}$$

$$B_b = b\frac{dP}{da} = \frac{B_a}{\alpha} \tag{16}$$

$$B_c = c\frac{dP}{da} = \frac{B_a}{\beta} \tag{17}$$

$$B_{relax} = \frac{\Lambda}{(1 + \alpha + \beta)^2} \tag{18}$$

where,

$$\Lambda = C_{11} + 2C_{12}\alpha + C_{22}\alpha^2 + 2C_{13}\beta + C_{33}\beta^2 + 2C_{23}\alpha\beta$$

Here,

$$\alpha = \frac{\{(C_{11} - C_{12})(C_{33} - C_{13})\} - \{(C_{23} - C_{13})(C_{11} - C_{13})\}}{\{(C_{33} - C_{13})(C_{22} - C_{12})\} - \{(C_{13} - C_{23})(C_{12} - C_{23})\}}$$

and,

$$\beta = \frac{\{(C_{22} - C_{12})(C_{11} - C_{13})\} - \{(C_{11} - C_{12})(C_{23} - C_{12})\}}{\{(C_{22} - C_{12})(C_{33} - C_{13})\} - \{(C_{12} - C_{23})(C_{13} - C_{23})\}}$$

**Table 4** The bulk modulus ($B_{relax}$ in GPa), bulk modulus along the crystallographic axes $a$, $b$, $c$ ($B_a$, $B_b$, $B_c$), $\alpha$ and $\beta$ of BeB$_2$C at the ground state.

| $B_{relax}$ | $B_a$ | $B_b$ | $B_c$ | $\alpha$ | $\beta$ | Ref. |
|---|---|---|---|---|---|---|
| 189.91 | 410.21 | 410.21 | 2563.84 | 1.00 | 0.16 | [This work] |

The phenomenon of elastic anisotropy constitutes a critical property attributable to the divergent bonding characteristics exhibited along various crystallographic orientations in crystalline solids. Consequently, it is imperative to compute the elastic anisotropy in structural intermetallic to elucidate these properties and to effectively identify mechanisms that enhance their longevity. In the realm of materials design, particularly concerning compounds characterized by layered structures, the investigation of elastic anisotropy represents a significant endeavor within the fields of engineering science and crystallography [42]. For BeB$_2$C, we have computed several anisotropy indices at various pressures, which we have displayed in Table 5. The most commonly used anisotropy indices are the universal anisotropy factors ($A^U$, $d_E$), which is applicable to all crystal systems regardless of symmetry. The shear anisotropy factors ($A_1$, $A_2$, and $A_3$), Equivalent Zener anisotropy factor ($A^{eq}$), the anisotropy indices for the bulk and shear moduli ($A_B$ and $A_G$) have calculated. Besides, we have calculated the compressibility ratio along $c$- and $a$-direction, $k_c/k_a$, of BeB$_2$C which is another important anisotropy parameter [43,44]. All the calculated values are presented in Table 5. The following relations are used to calculate the anisotropy indices:

$$A = \frac{2C_{44}}{C_{11} - C_{12}} \tag{19}$$

$$A_1 = \frac{4C_{44}}{C_{11} + C_{33} - 2C_{13}} \tag{20}$$

$$A_2 = \frac{4C_{55}}{C_{22} + C_{33} - 2C_{23}} \tag{21}$$

$$A_3 = \frac{4C_{66}}{C_{11} + C_{22} - 2C_{12}} \tag{22}$$

$$A_B = \frac{B_V - B_R}{B_V + B_R} \quad and \quad A_G = \frac{G_V - G_R}{G_V + G_R} \tag{23}$$

$$A^U = \frac{B_V}{B_R} + 5\frac{G_R}{G_R} - 6 \geq 0 \tag{24}$$

$$\frac{k_c}{k_a} = \frac{C_{11} + C_{12} - 2C_{13}}{C_{33} - C_{13}} \tag{25}$$

$$d_E = \sqrt{A^U + 6} \tag{26}$$

$$A^{eq} = \left(1 + \frac{5}{12}A^U\right) + \sqrt{\left(1 + \frac{5}{12}A^U\right)^2 - 1} \tag{27}$$

From the Table 5, we can see that, the value of shear anisotropy ($A_1$, $A_2$, and $A_3$) is unity, which indicates the isotropic nature of the material with respect to shape deformation.

**Table 5** Calculated indices of elastic anisotropy and compressibility ratio of BeB$_2$C.

| $A$ | $A_1$ | $A_2$ | $A_3$ | $A_B$ | $A_G$ | $A_U$ | $d_E$ | $A^{eq}$ | $k_c/k_a$ | Ref. |
|------|------|------|------|------|------|------|------|------|------|------|
| 1.55 | 0.57 | 0.57 | 1.00 | 0.04 | 0.12 | 1.41 | 2.72 | 2.59 | 0.51 | [This work] |

The overall anisotropy index ($A$ = 1.55) suggests that the elastic properties of material differ across different crystallographic directions, influencing its mechanical behavior under stress. The individual anisotropy factors, $A_1$ and $A_2$, are both 0.57, while $A_3$ = 1.00, indicating uniform elastic behavior along one direction and moderate variation in others. $A$ = 1.00 means a completely isotropic material, whereas a value smaller or larger than unity indicates the degree of elastic anisotropy [45]. Anisotropy in the bulk and shear moduli is minimal, with $A_B$ = 0.04 and $A_G$ = 0.12, reflecting consistent compressive and shear behavior across different planes and directions. The universal anisotropy index ($A_U$) of 1.41 further supports the presence of some anisotropy, though the material remains closer to isotropy than highly anisotropic materials. The compressibility ratios, $d_E$ = 2.72 and $A^{eq}$ = 2.59, show that BeB$_2$C compresses more easily along specific directions, with the $k_c/k_a$ value of 0.51 indicating that compressibility is approximately half as much in one axis compared to the other. These findings highlight the directional dependence BeB$_2$C in both elastic and compressive properties, which is crucial for understanding its performance under high-pressure conditions.

The two- and three-dimensional dependences of the Young's modulus, Linear compressibility, Shear modulus and Poisson's ratio were calculated for BeB$_2$C using the ELATE program and presented in Figure 2. The plots for isotropic materials have spherical shape. The amount of deviation in Figure 2 indicates the degree of anisotropy. As can be clearly seen from all planes in Figure 2, the system is anisotropic [46]. The curves plotted in green and blue represent the minimum and maximum values for the respective parameters. It is apparent that Young's modulus and linear compressibility of BeB$_2$C exhibit isotropy within the *xy*-plane, while displaying anisotropic characteristics in all other planes. Given the lack of both experimental and theoretical data pertaining to BeB$_2$C, we posit that our results regarding the examined properties will provide a valuable reference for forthcoming experimental and theoretical investigations.

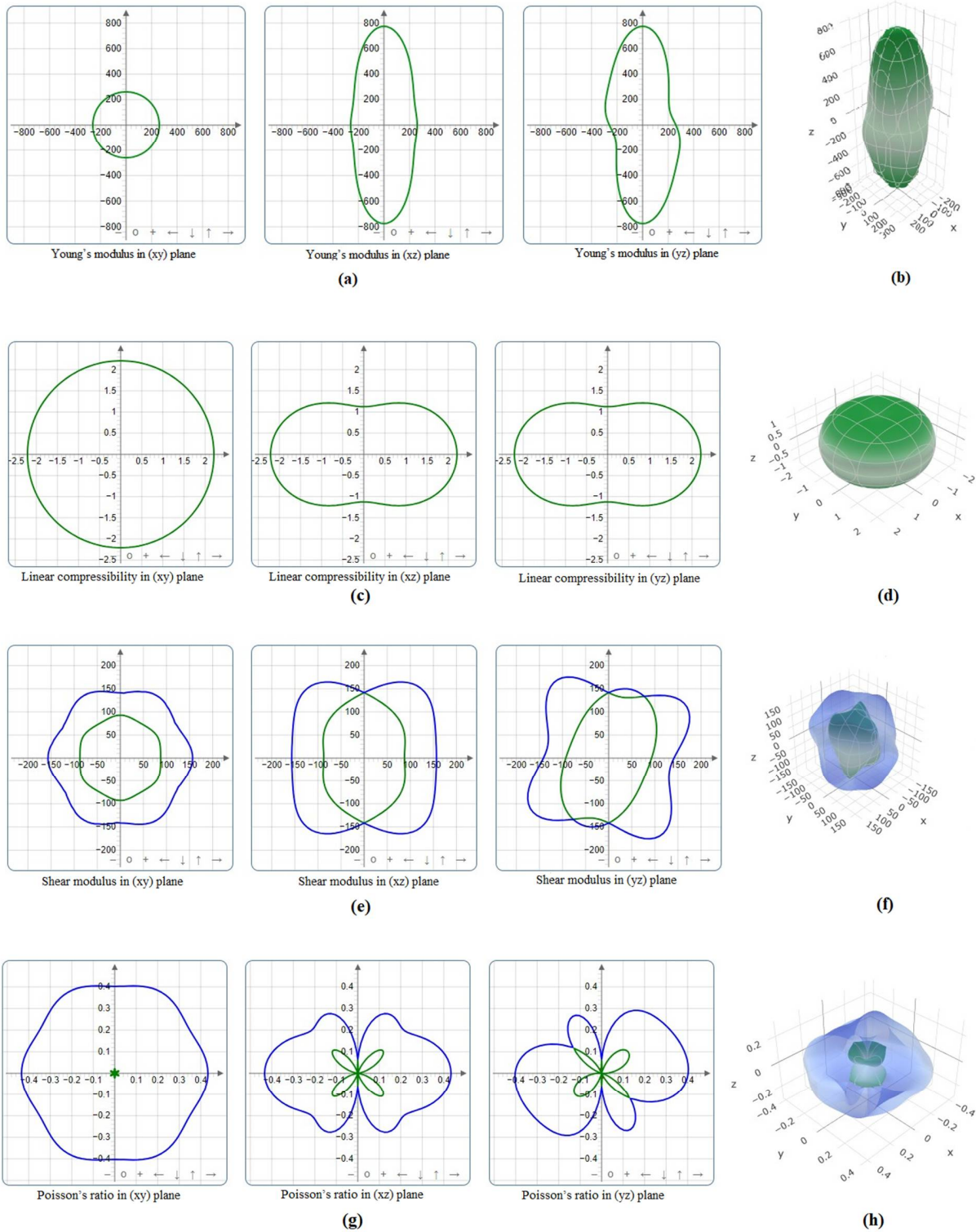

**Figure 2** Two-dimensional (2D) directional dependences of (a) Young's modulus, (c) Linear Compressibility, (e) Shear Modulus and (g) Poisson's ratio; Three-dimensional (3D) directional dependences of (b) Young's modulus, (d) Linear Compressibility, (f) Shear Modulus and (h) Poisson's ratio of BeB$_2$C.

### 3.4 Electronic properties

*(a) Electronic band structure*

In order to better understand electrical, magnetic, and optical properties of solid materials, band structures are utilized to illustrate the permitted electronic energy levels. In a crystalline substance, the energy of the atomic orbitals is represented in two dimensions by a band structure.

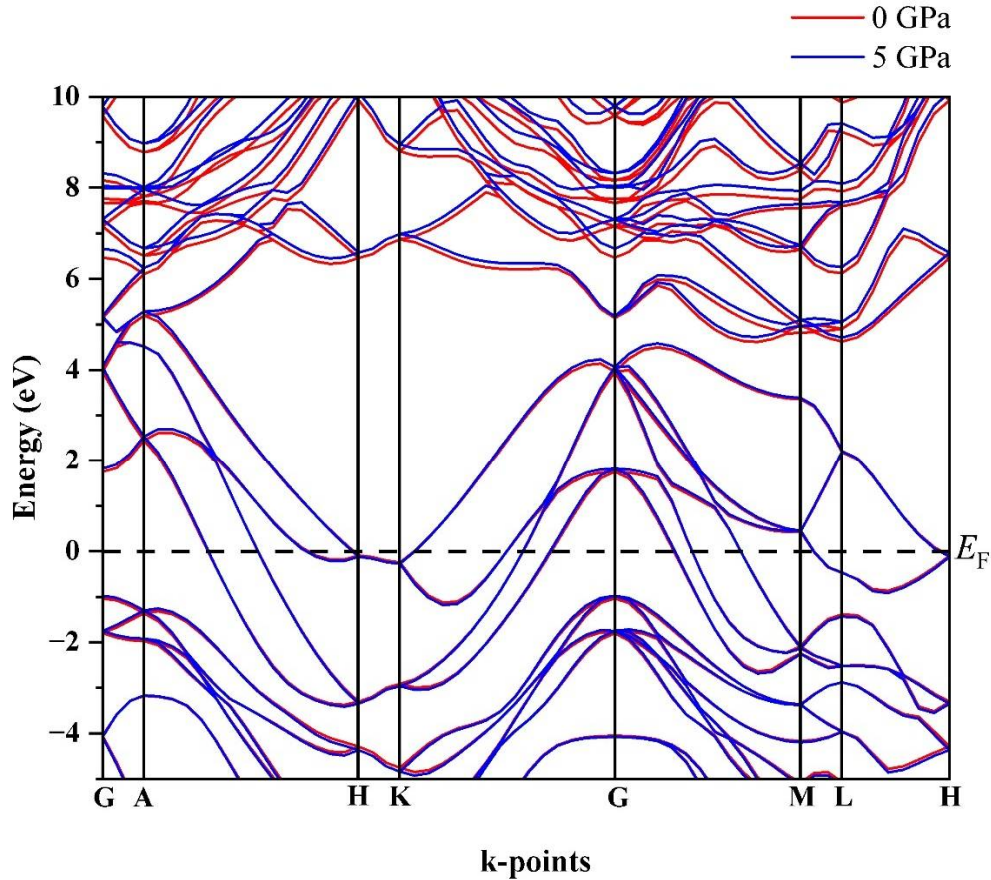

**Figure 3** The band structure of BeB$_2$C along the high symmetry directions of the *k*-space within the first Brillouin zone at 0 GPa (Red line) and 1 GPa (Blue line).

The metallic, semi-metallic, and insulating properties of a material can be rapidly determined using a band structure map. Additionally, the curvature of bands can reflect the carrier mobility in those bands. The band structures with LDA (CA-PZ) functional along high-symmetry direction in Brillouin zone under pressure of *R*3m-BeB$_2$C are shown in Figure 3. The blue dash line signifies the Fermi energy level ($E_F$), which is defined as the highest energy level occupied by the valence electrons at 0 K. The vertical pink lines in the band structure represent the points in the direction of high symmetry. In the figure, *G, A, H, K, G, M, L, H* were used to describe the high symmetry points.

As can be seen from the band structure, the valence and conduction bands overlap considerably, and there is no band gap at the Fermi level, which indicates that BeB$_2$C exhibits metallic property. In Figure 3, we see that pressure shifts the band energy to a higher value. This effect is small in the valence bands but significant in the high energy conduction bands. The band widths are quite insensitive to applied pressure in the pressure range considered. The bands are quite dispersive and the effective mass of the charge carriers is expected to be low [47]. Overall, the band structure suggests that BeB$_2$C has a complex electronic structure with multiple bands crossing the Fermi level. From the curvatures of the bands crossing the Fermi level, we expect both electron and hole Fermi sheets in the Fermi surface of this compound.

*(b) Density of States*

The number of electronic states per unit energy per unit volume that can be occupied by electrons in a material is described by the electronic density of states (EDOS). The EDOS and the electron distribution, governed by the Fermi-Dirac statistics, determine the total number of electrons that can occupy any particular energy level. The calculated total density of states (TDOS) and atomic orbital specific partial density of states (PDOS) for BeB$_2$C are shown in Figure 4 under different pressures along the high symmetry points in the Brillouin zone (BZ).

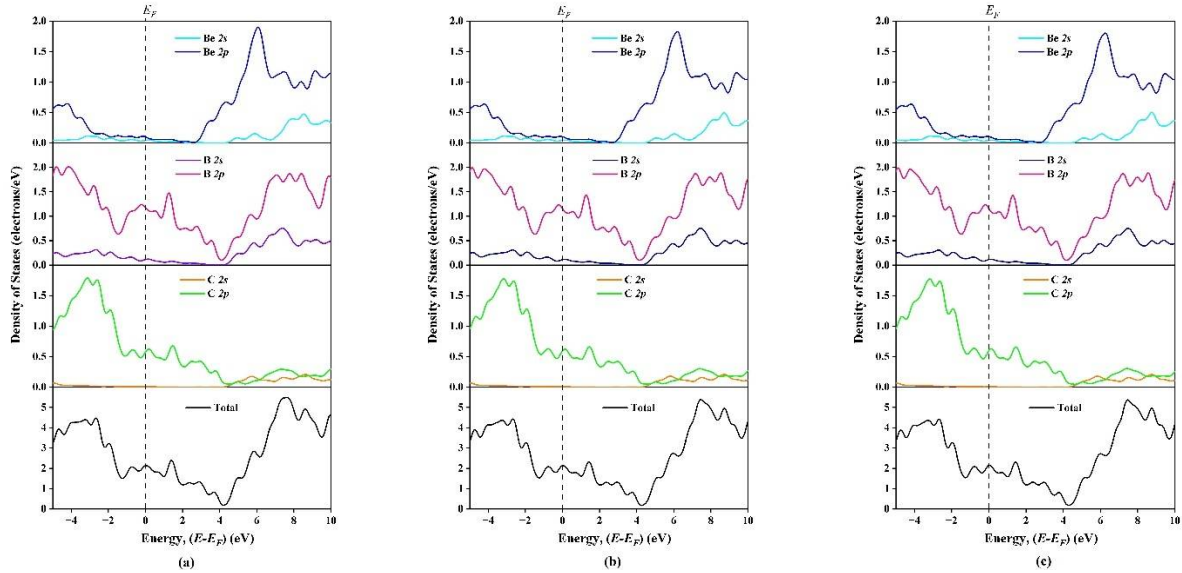

**Figure 4** Total and partial density of states of BeB$_2$C at (a) 0 GPa, (b) 4 GPa and (c) 5 Gpa.

The BeB$_2$C electronic density of states (DOS) and atom resolved partial density of states (PDOS) as a function of energy ($E$-$E_F$) are shown in Figure 4 at various pressures. The dashed vertical line at 0 eV represents the Fermi level ($E_F$). Positive values represent energy states above the Fermi level (unoccupied states), and negative values represent energy states below the Fermi level (occupied states). The density of electronic states, expressed in electrons per electron volt (eV), is displayed on the *y*-axis. From Figure 4, it can be seen that TDOS for BeB$_2$C has a non-zero value at the Fermi level at 0 GPa, which means that there are available electronic states at the Fermi level that can be occupied by electrons demonstrating that BeB$_2$C is a metal. The non-zero TDOS at the Fermi level across all pressures suggests that the material retains its metallic properties under pressure. The material electronic structure of the material is stable under the applied pressures, as evidenced by the general similarity in the shape of the DOS curves at various pressures. From Figure 4, it can be seen that, in case of the valence band in the lower energy

−5 eV to 0 eV, the contribution of Be-2$p$, B-2$s$, B-2$p$, and C-2$p$ states are comparatively significant. It is also observed from Figure 4 that the contribution of the Be-2$s$ and C-2$s$ orbitals to the valence band is minimal. The 2$p$ states of Be show significant density of states around 5-9 eV above the Fermi level, suggesting these are the major contributors to the conduction band. Boron 2$p$ states are also significant, particularly between 1-5 eV above the Fermi level, contributing to the material's conduction. The C-2$p$ states contribute primarily at energies below the Fermi level and slightly above, influencing the valence band and play prominent role in chemical bonding. At the Fermi level the B-2$p$ orbital contribution is dominant. The presence of significant DOS at the Fermi level suggests that the material exhibits metallic behavior. Accordingly, the occupancy of the bonding orbitals raises the strength of the bond and thereby increasing the bulk moduli [48] of BeB$_2$C.

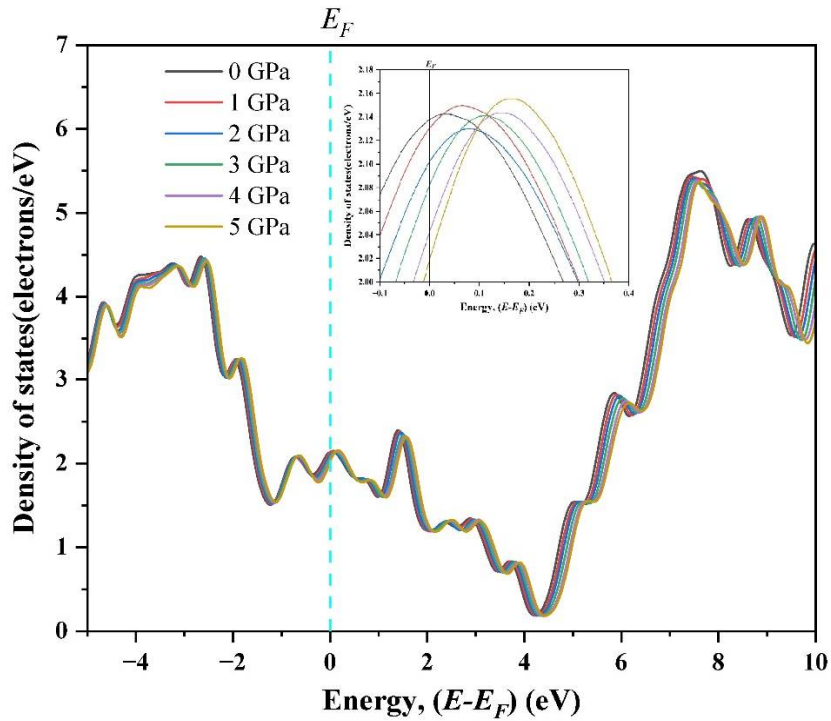

**Figure 5** Comparison of the total density of states of BeB$_2$C at 0 GPa, 1 GPa, 2 GPa, 3 GPa, 4 GPa and 5 GPa.

From Figure 5 it can be seen that, when the external pressure is increased, the peaks in the conduction band shift toward the higher energy, showing that the covalency of BeB$_2$C increases with pressure. This is the main reason of the reduction in TDOS under pressure at the Fermi level. We have already noted that the elastic constant and modulus of BeB$_2$C increase with increasing covalency [49].

**Table 6** Calculated total density of states at the Fermi level, N($E_F$) in electrons/eV-unit cell, Coulomb pseudopotential, $\mu^*$ of BeB$_2$C under different pressures.

| P (GPa) | N($E_F$) | $\mu^*$ | Ref. |
|---|---|---|---|
| 0 | 2.139 | 0.177 | [This work] |
| 1 | 2.133 | 0.177 | |
| 2 | 2.105 | 0.176 | |
| 3 | 2.086 | 0.176 | |
| 4 | 2.048 | 0.175 | |
| 5 | 2.025 | 0.174 | |

The total density of states, N($E_F$) per unit cell at the Fermi level for BeB$_2$C, along with the Coulomb pseudopotential $\mu^*$ for BeB$_2$C under varying pressures ranging from 0 GPa to 5 GPa were listed in Table 6. In this case, the Fermi level represents the highest occupied state at absolute zero temperature, and N($E_F$) provides insight into the number of electronic states available at Fermi level. The distribution of PDOS reflects the localized bonding character of Boron and Carbon (especially from $sp^2$ hybridization) in the valence band and the delocalized conduction states where Beryllium influence the electronic conductivity of the material. The Coulomb pseudopotential, $\mu^*$, is an important parameter in superconductivity studies, reflecting the electron-electron repulsion in the material. The repulsive Coulomb pseudopotential, $\mu^*$, can be determined directly by the TDOS at the Fermi level. We have calculated the $\mu^*$ of BeB$_2$C materials by [50]:

$$\mu^* = \frac{0.26 \, N(E_F)}{1 + N(E_F)} \tag{28}$$

Analyzing the table, it is evident that as pressure increases from 0 GPa to 5 GPa, there is a gradual decrease in the total density of states N($E_F$) at the Fermi level. For instance, N($E_F$) decreases from 2.139 at 0 GPa to 2.025 at 5 GPa. This reduction suggests that the number of electronic states available at the Fermi level diminishes as the system is subjected to higher pressure, which is associated with pressure-induced modifications in the electronic structure, such as band broadening or shifting of energy bands. Furthermore, the Coulomb pseudopotential $\mu^*$ also shows a slight decrease, from 0.177 at 0 GPa to 0.174 at 5 GPa. This subtle variation in $\mu^*$ could indicate a minor reduction in the electron-electron repulsion with increasing pressure, which might weakly affect the superconducting properties of the material. Overall, these pressure-dependent changes in N($E_F$) and $\mu^*$ reflect the complex interplay between electronic structure and external pressure [51] in BeB$_2$C, suggesting possible implications for its conductivity and superconductivity under different pressure conditions.

*(c) Fermi surface*

The Fermi surface in momentum space of a crystalline solid represents a paradigm of quantum solid state physics that separates the occupied electron states from the unoccupied ones at zero temperature. Its shape is dictated by quantum mechanics, by the Fermi-Dirac statistics for electrons and by the character of Bloch states in solids. The shape of the Fermi surface determines most physical observables in one way or another. These factors have led to investigations of the Fermi surface of solids for various materials using various methodologies [52-54].

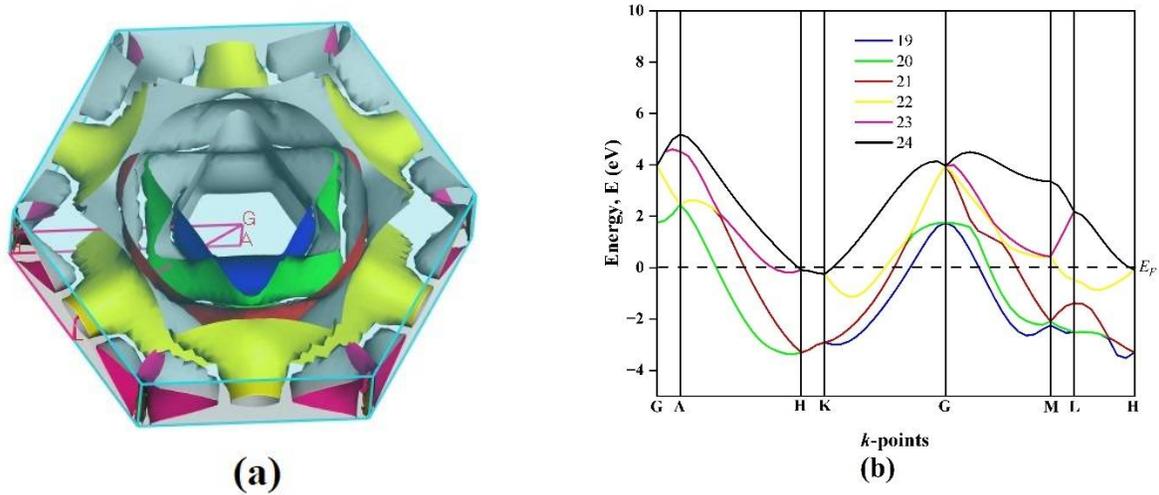

**Figure 6** (a) Calculated six bands crossing Fermi surface of BeB$_2$C along $\Gamma$-A direction, (b) Associated bands (19, 20, 21, 22, 23, 24) cross the Fermi level in the band structure at the ground state.

Fermi surfaces (FS) in the first Brillouin zone of the six bands crossing the Fermi energy are presented in Figure 6(a). The band crossing line in the Fermi level was indicated with a color which is shown in Figure 6(b). From Figure 6(b) it can be seen that, all bands (all are conduction bands) cross the Fermi energy multiple times along the high symmetry BZ region.

We show FS view from the top for each corresponding band in Figure 7. From Figure 6 it can be seen that, the Fermi surface center is comprised of six hexagon-shaped, both electron and hole-like sheets with varying topologies, oriented along the $\Gamma$-A direction. The 19, 20, 21 bands in Figure 7 (a), (b), (c) consisting of hexagonal cylindrical like shape or electron-like sheets enclosing the middle of the Brillouin zone has electronic character. These are seen to have bulged surfaces. These bulging shapes indicate electron-like pockets, as they enclose regions of higher energy states in the Brillouin zone. The fourth sheet has complex lateral funnel type shape with hexagonal cross-section. The hexagonal hourglass like shape of the FS of band 23 and 24 band are hole-like. The weak crossing leads to a small inwardly curved surfaces or concavities near the outer edges of the Brillouin zone (*M*, *K*, *H*) representing hole-like pockets [55]. The pressure effect on the FS is negligible, so we only performed this calculation at ground state (0 GPa).

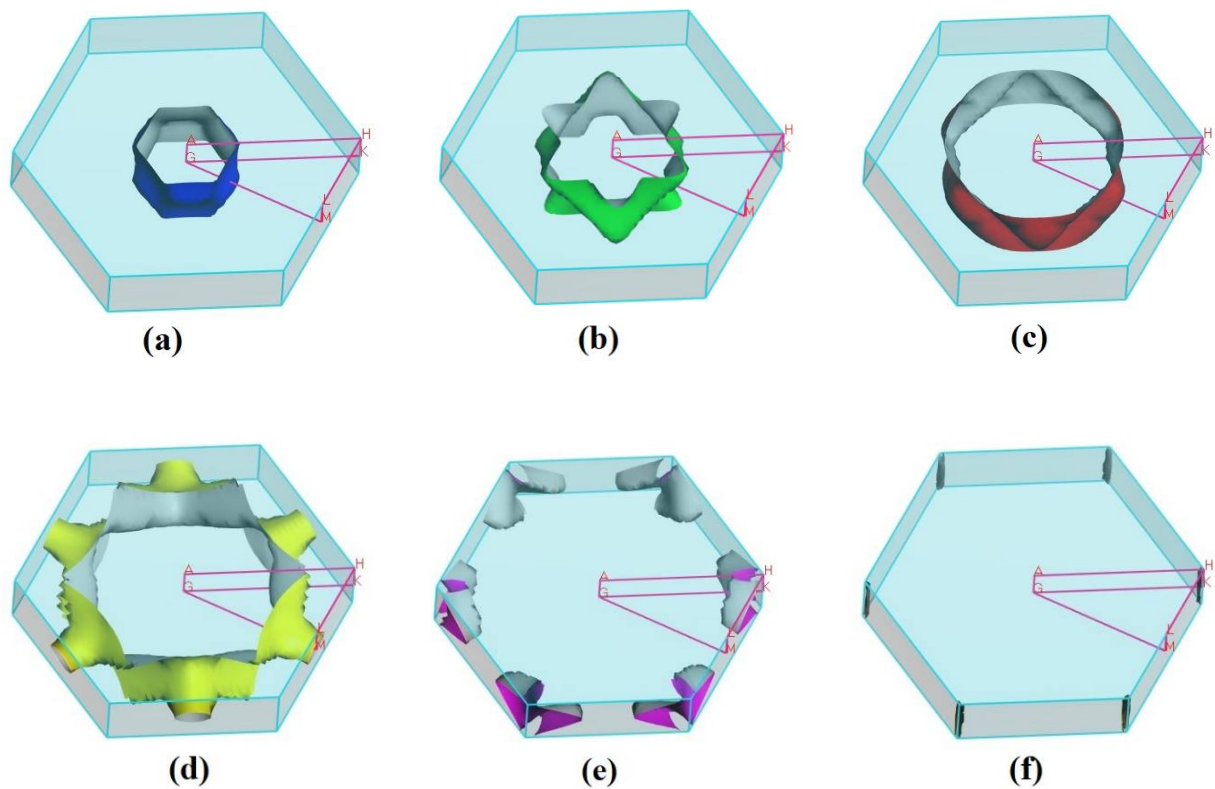

**Figure 7** The Fermi surfaces of BeB$_2$C for band (a) 19, (b) 20, (c) 21, (d) 22, (e) 23 and (f) 24 with hexagonal Brillouin zone along $\Gamma$-A direction.

### 3.5 Bond population analysis and theoretical bond hardness

We have conducted a comprehensive analysis employing both Mulliken bond population [20] analysis and Hirshfeld population analysis [56] to gain an in-depth understanding of the bonding characteristics (ionic, covalent, and metallic) within the BeB$_2$C compound, as well as the effective valence of the constituent atoms present in this compound. The Mulliken bond population analysis offers a quantitative framework for the determination of bond characteristics, elucidating significant charge transfer between the atoms, which suggests a mixed character of bonding. Conversely, the Hirshfeld population analysis (HPA), which provides a more localized perspective by distributing the electron density into regions associated with individual atoms, affords a more sophisticated understanding of the atomic contributions to the bonding. The outcomes of these analyses are systematically presented in Table 7. The comparatively low value of the charge spilling parameter (0.40%) signifies that the results obtained from Mulliken population analysis (MPA) are of high quality. Mulliken charges are computed through the assessment of the electron population attributable to each atom as defined within the basis functions [57]. According to MPA, the beryllium (Be) atoms primarily contribute via *s*-orbital electrons, yielding a Mulliken population of 2.13, whereas the contributions from *p*-orbitals are negligible. In contrast, boron (B) and carbon (C) atoms exhibit substantial *p*-orbital contributions, which reflect their involvement in forming covalent bonds within the molecular structure. The Mulliken charge analysis indicates that Be atoms exhibit a positive charge of +0.83 e, implying their role as electron donors, whereas B and C atoms display slightly negative charges, suggesting their function as electron acceptors. Moreover, from the HPA perspective, Be demonstrates an effective valence of +1.89, which is in proximity to its formal ionic charge of +2, while B and C atoms present values around −1.94 and −0.18, respectively. These findings underscore that the BeB$_2$C compound manifests a mixed bonding nature, with ionic character emanating from the electron donation by Be and covalent character arising from interactions between B and C atoms. The analyses reveal that, although MPA and HPA may produce differing numerical values due to their distinct partitioning methodologies, the trends in charge distribution and valence states among the Be, B, and C atoms exhibit alignment between the two analytical approaches, thereby suggesting a qualitative concordance in the depiction of electron transfer and bonding characteristics within the compound.

**Table 7** Charge Spilling parameter (%), orbital charge (electron), atomic Mulliken charge (electron), effective valance (Mulliken & Hirshfeld) (electron) of BeB$_2$C in the ground state.

| Charge spilling | Atomic species | Mulliken Atomic Population-orbitals | | | Mulliken charge | Formal ionic charge | Effective valence (Mulliken) | Hirshfeld Charge | Effective valence (Hirshfeld) |
|---|---|---|---|---|---|---|---|---|---|
| | | s | p | Total | | | | | |
| 0.40 | Be | 2.13 | 1.04 | 3.17 | 0.83 | +2 | 1.17 | 0.11 | 1.89 |
| | Be | 2.13 | 1.04 | 3.17 | 0.83 | +2 | 1.17 | 0.11 | 1.89 |
| | Be | 2.13 | 1.04 | 3.17 | 0.83 | +2 | 1.17 | 0.11 | 1.89 |
| | B | 0.84 | 2.25 | 3.09 | −0.09 | −2 | 1.91 | 0.02 | 1.98 |
| | B | 0.79 | 2.22 | 3.01 | −0.01 | −2 | 1.99 | 0.06 | 1.94 |
| | B | 0.84 | 2.25 | 3.09 | −0.09 | −2 | 1.91 | 0.02 | 1.98 |
| | B | 0.79 | 2.22 | 3.01 | −0.01 | −2 | 1.99 | 0.06 | 1.94 |
| | B | 0.84 | 2.25 | 3.09 | −0.09 | −2 | 1.91 | 0.02 | 1.98 |
| | B | 0.79 | 2.22 | 3.01 | −0.01 | −2 | 1.99 | 0.06 | 1.94 |
| | C | 1.26 | 3.47 | 4.73 | −0.73 | 0 | −0.73 | −0.18 | −0.18 |
| | C | 1.26 | 3.47 | 4.73 | −0.73 | 0 | −0.73 | −0.18 | −0.18 |
| | C | 1.26 | 3.47 | 4.73 | −0.73 | 0 | −0.73 | −0.18 | −0.18 |

The hardness of BeB$_2$C is strongly linked to its bonding structure, where the presence of strong covalent/ionic bonds suggests high hardness. Hardness, defined as the resistance to both elastic and plastic deformation, is particularly intrinsic to covalent crystals and is determined by the resistance of individual bonds to deformation, quantified by the energy gap and bond density. Valence electron density plays a key role in determining the number of bonds per unit area, directly impacting hardness [58]. A semi-empirical method can be used to evaluate hardness in covalent crystals, incorporating corrections for compounds with delocalized metallic bonding. Gou *et al*. [59] proposed an equation that accounts for this metallic bonding correction when estimating hardness in such materials. The positive (+) and negative (−) values of overlap population denote the existence of bonding-type and anti-bonding-type interactions among the atoms, respectively. An overlap population approaching zero signifies the absence of substantial interaction between the electronic populations of the two bonding atoms. Therefore, BeB$_2$C, with its strong covalent character and possible metallic bonding features, holds the potential for high hardness due to the significant resistance from both its covalent and delocalized bonds. Thus, the hardness can be calculated using the following relations [60,61]:

$$H_v^\mu = 740(P^\mu - P^{\mu\prime})(v_b^\mu)^{-5/3} \tag{29}$$

$$H_v = \left[\prod (H_v^{\mu n^\mu})\right]^{1/\Sigma n^\mu} \tag{30}$$

where,

$$P^{\mu\prime} = \frac{n_{free}}{V_0} \tag{31}$$

$$n_{free} = \int_{E_P}^{E_F} N(E)\, dE \tag{32}$$

$$v_b^\mu = \frac{(d^\mu)^3}{\Sigma_v[(d^\mu)^3 n_b^v]} \tag{33}$$

$P^\mu$ is the Mulliken overlap population, $v_b^\mu$ is the bond volume, $n^\mu$ is the number of $\mu$-type bond, $d^\mu$ is the bond length, and $n_b^v$ is the total number of $v$-type bond number per unit volume. $P^{\mu\prime}$ is the metallic population and is calculated from the cell volume, $V$, and the number of free electrons in a cell, $n_{free}$. In Table 8, we present the bond parameters and hardness for BeB$_2$C at zero pressure. It was found that the B–C bonds have shorter bond lengths and higher hardness values of 115.686 GPa and 111.074 GPa, which greatly surpass the hardness of B–B bond of 9.782 GPa. This indicates that the B–C bonds are more resistant to deformation and contribute significantly to the mechanical stability of the compound. Thus, while B–B bonds share more electrons and show the higher degree of covalency, B–C bond show more resistance to deformation. Unfortunately, as far as we know, there are no experimental and theoretical data available related to the hardness of BeB$_2$C in the literature for comparison.

**Table 8** Calculated bond overlap population of $\mu$-type bond ($P^\mu$), total number of bonds ($n^\mu$) and bond lengths ($d^\mu$ in Å), hardness of $\mu$-type bond ($H_v^\mu$), and bond hardness ($H_v$ in GPa) for BeB$_2$C.

| Bond | $P^\mu$ | $n^\mu$ | $d^\mu$ | $P^{\mu\prime}$ | $v_b^\mu$ | $H_v^\mu$ | $H_v$ | Ref. |
|---|---|---|---|---|---|---|---|---|
| B-C | 0.88 | 6 | 1.48300 | 0.006 | 2.809 | 115.686 | | |
| B-C | 0.85 | | 1.48471 | 0.006 | 2.818 | 111.074 | | |
| Be-C | 0.98 | 3 | 1.77901 | 0.006 | 4.848 | 51.897 | 34.3 | [This work] |
| Be-B | 0.11 | 6 | 1.88517 | 0.006 | 5.769 | 4.147 | | |
| Be-B | 0.33 | | 1.99305 | 0.006 | 6.139 | 60.191 | | |
| B-B | 1.68 | 3 | 1.92460 | 0.006 | 6.817 | 9.782 | | |

It is noteworthy to observe that, based on the concept of effective valence, it is occasionally challenging to quantify the precise degree of ionicity and covalency. The overlap population serves as a qualitative indicator of the trend pertaining to bond ionicity and covalency. A significantly elevated positive overlap population suggests a diminished level of ionicity or an enhanced level of covalency (for instance, the sharing of electrons facilitated by the overlapping of atomic orbitals between atoms) within the context of chemical bonding.

From Table 8, it is evident that metallic bonding possesses a soft characteristic and contributes minimally to the overall hardness of a material. The metallic concentration for BeB$_2$C is significantly low. The computed values pertaining to the theoretical hardness of BeB$_2$C are presented in Table 8. The elevated bond hardness of BeB$_2$C, determined through bond hardness calculations, corroborates the conclusions derived from the examination of elastic constants and moduli pertaining to the ternary boride. Based on the Mulliken overlap population analysis, it is also found that BeB$_2$C is a significantly hard material.

### 3.6 Electron density difference

To visualize the nature of the bond character and to explain the charge transfer and bonding properties of BeB$_2$C we have calculated and plotted their valence electron density difference maps in the (001) and (111) planes as illustrated in Figure 8.

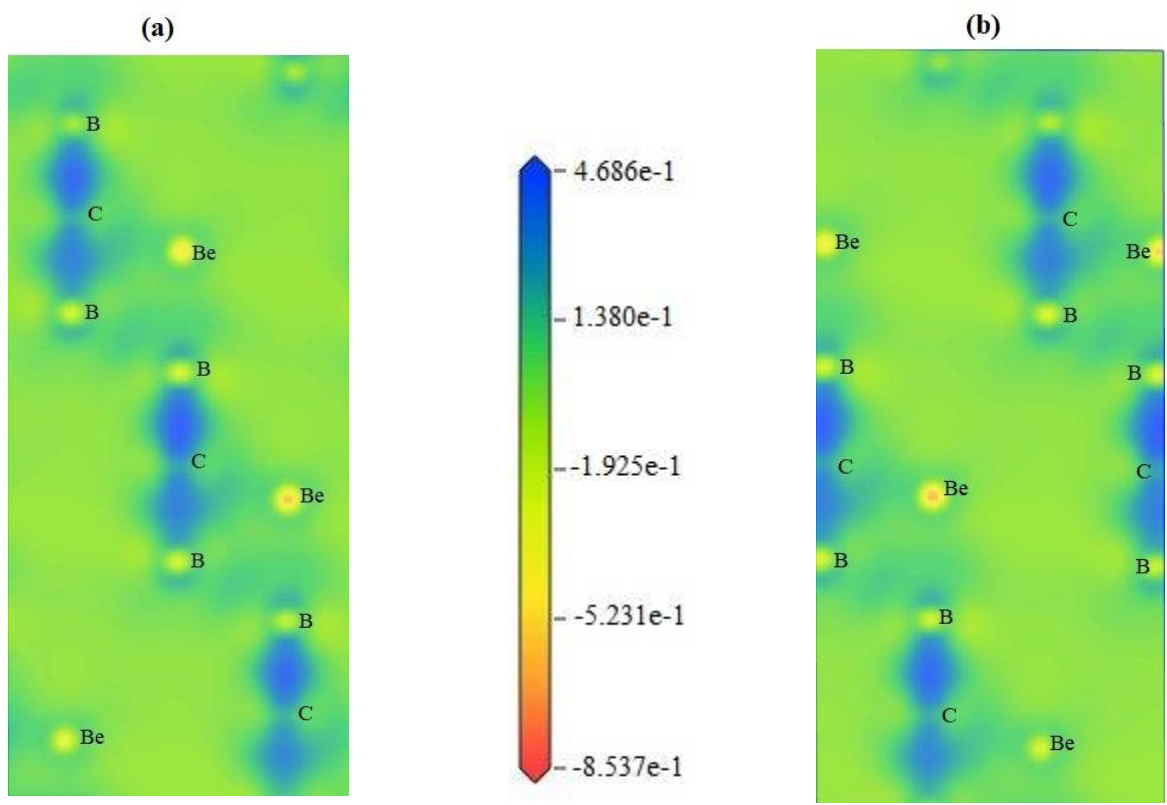

**Figure 8** Calculated charge density difference of the titled BeB$_2$C along (a) (001) and (b) (111) planes with adjacent scale showing the charge value (in e/ Å$^3$).

The total electron density for each plane is shown by the color scale in the middle of the charge density maps; for BeB$_2$C, blue denotes a high charge density and red denotes a low charge density. Figure 8 illustrates the calculated charge density difference for BeB$_2$C compound along two distinct crystallographic planes, specifically the (a) (001) and (b) (111) planes. The color scale adjacent to the images quantifies the charge density difference in terms of electronic charge units, where dark red, indicating areas of electron depletion and dark blue, indicating regions of electron accumulation.

In both planes, a notable concentration of charge density (electron localization) is visible around the beryllium (Be) and carbon (C) atoms, which are marked within the structure. Beryllium appears to exhibit significant electron depletion, shown by the lower charge density values in its vicinity, as indicated by the greenish-yellow regions, especially when compared to the boron (B) atoms. In contrast, the charge around carbon and boron atoms exhibits a more varied distribution, with boron displaying regions of both charge depletion and accumulation. It can also be seen from Figure 8, that, B and C atoms in BeB$_2$C form bonds where B atoms lose electrons and C atoms get electrons. The charge distribution patterns suggest that the bonding characteristics in the BeB$_2$C material are not homogeneous, with differential electron density surrounding each type of atom. This variation likely points to the mixed covalent and ionic nature of bonding within the material, where beryllium and carbon may be involved in stronger covalent interactions, while boron atoms experience varied electronic environments depending on their position. According to the Mulliken charge analysis (Table 7), beryllium (Be) atoms have a net positive charge (+0.83), indicating electron depletion, which agrees with the greenish-yellow regions around Be in the charge density map, suggesting lower electron localization. Similarly, carbon (C) atoms have a negative Mulliken charge (−0.73), which matches the observed electron accumulation (blue regions) in the vicinity of carbon atoms in the charge density map. The boron (B) atoms show almost neutral to slightly negative Mulliken charges (−0.01 to −0.09), indicating a more balanced electron distribution, which is consistent with the intermediate charge density values seen around boron in the visual data.

Based on the both of Mulliken charge analysis and the charge density maps, BeB$_2$C demonstrates mixed bonding characteristics where carbon forms predominantly covalent bonds with boron, while beryllium participates in both ionic and weaker covalent interactions, resulting in a complex bonding network essential for its structural stability, which could impact on potential applications in advanced materials science.

### 3.7. Thermophysical properties

A number of technologically important thermophysical properties of BeB$_2$C are investigated in this section. These parameters are linked to the elastic constants and moduli.

(a) *Debye temperature* ($\theta_D$)

The Debye temperature $\theta_D$ is an important fundamental parameter in solids, which has relationships with some very important physical properties such as specific heat, melting temperature, elastic constants and thermal conductivity. Using the calculated elastic constants data, we can estimate the Debye temperature, $\theta_D$, using the following equation [62]:

$$\theta_D = \frac{h}{k_B}\left[\frac{3n}{4\pi}\left(\frac{N_A\rho}{M}\right)\right]^{\frac{1}{3}} v_m \qquad (34)$$

where, $h$ is the Planck's constant, $k_B$ is the Boltzmann's constant, $N_A$ is the Avogadro's number, $\rho$ is mass density, $M$ is the molecular weight, $v_m$ is the mean sound velocity and $n$ denotes the number of atoms in the cell. The mean sound velocity, $v_m$ in the crystal is calculated from,

$$v_m = \left[\frac{1}{3}\left(\frac{2}{v_t^3} + \frac{1}{v_l^3}\right)\right]^{-\frac{1}{3}} \tag{35}$$

where, $v_l$ and $v_t$ represent the longitudinal and transverse modes of sound velocities. These can be calculated from the bulk modulus, $B$ and shear modulus, $G$:

$$v_t = \sqrt{\frac{G}{\rho}} \quad and \quad v_l = \sqrt{\frac{3B + 4G}{3\rho}} \tag{36}$$

From Table 9 it can be seen that the Debye temperature decreases sharply from 1256.37 K at 0.0 GPa to 1149.51 K at 3.0 GPa, indicating a reduction in the vibrational energy of the lattice under moderate pressures. However, after this point, it starts to rise again, reaching 1261.99 K at 5.0 GPa. Sound velocities, both transverse ($v_t$) and longitudinal ($v_l$), exhibit small fluctuations across the pressure range, with a slight dip around 3.0 GPa corresponding to the drop in $\theta_D$. This correlation suggests that reduced sound velocities at moderate pressures reflect a softening ability of the material to propagate phonons. As the velocities recover at higher pressures, the increase in $\theta_D$ indicates that the material becomes stiffer, enhancing its phonon propagation and thermal characteristics [63]. The mean sound velocity ($v_m$) follows a similar pattern, supporting this relationship between sound propagation and lattice vibrations under pressure.

(b) *Melting temperature* ($T_m$)

Melting temperature is the temperature of a substance at which a solid turns into liquid, and is an important parameter for materials. It was predicted by Fine *et al.* [64] using the elastic constants of the crystal:

$$T_m = 354 + 1.5(2C_{11} + C_{33}) \tag{37}$$

From Table 9 it can be seen that, the melting temperature varies with pressure in a non-linear pattern. It initially increases from 2431.14 K at 0.0 GPa to a peak of 2492.37 K at 2.0 GPa, after which it decreases at 3.0 GPa to 2346.69 K, and then rises again up to 2469.48 K at 5.0 GPa. This behavior reflects the complex interplay between atomic bonding and structural rearrangements under pressure. First-principles calculations often predict such trends, as pressure alters atomic distances and bonding strength, affecting the thermal stability and phase transition points of materials. For BeB$_2$C, the high Debye temperature values (ranging from 1149.51 K to 1261.99 K) suggest that it has a high melting temperature, which is characteristic of ceramic-like compounds with strong covalent bonding between their constituent atoms (beryllium, boron, and carbon). As pressure increases, the rising Debye temperature at higher pressures might also indicate that the melting point of BeB$_2$C increases, as the material becomes more thermally stable and less prone to softening at elevated temperatures. Thus, analyzing both $\theta_D$ and $T_m$ under pressure yields a consistent explanation: the increase in bond strength under higher pressures results in both higher vibrational frequencies (higher $\theta_D$) and greater thermal stability (higher $T_m$) [65].

The calculated Debye temperature, density of the unit cell ($\rho$), and sound velocities ($v_t$, $v_l$ and $v_m$) of BeB$_2$C are shown in Table 9. There are no previously reported data to compare to our results.

**Table 9** Density ($\rho$ in gm/cm$^3$), transverse, longitudinal, mean velocities of sound (all in km/sec), Debye temperature ($\theta_D$ in K) for BeB$_2$C under different pressures ($P$ in GPa).

| P | $\rho$ | $v_t$ | $v_l$ | $v_m$ | $\theta_D$ | $T_m$ | Ref. |
|---|---|---|---|---|---|---|---|
| 0.0 | 28.94 | 6.99 | 11.34 | 7.72 | 1256.37 | 2431.14 | |
| 1.0 | 29.10 | 6.99 | 11.40 | 7.71 | 1258.10 | 2461.18 | |
| 2.0 | 29.25 | 7.03 | 11.48 | 7.76 | 1267.88 | 2492.37 | [This work] |
| 3.0 | 29.41 | 6.32 | 10.97 | 7.02 | 1149.51 | 2346.69 | |
| 4.0 | 29.57 | 6.32 | 11.03 | 7.02 | 1151.39 | 2364.93 | |
| 5.0 | 29.74 | 6.96 | 11.31 | 7.68 | 1261.99 | 2469.48 | |

(c) *Minimum thermal conductivity* ($k_{min}$)

The thermal conductivity can be evaluated by using the average sound velocity and the formula is expressed as [66]:

$$\kappa_{min} = k_B v_m \left(\frac{nN_A\rho}{M}\right)^{\frac{2}{3}} \qquad (38)$$

The minimum thermal conductivity ($k_{min}$) of BeB$_2$C under pressure, as shown in Table 10, reflects a complex interplay between phonon transport, lattice dynamics, and structural changes. The minimum thermal conductivity generally increases with pressure, except for a noticeable drop at 3.0 GPa and 4.0 GPa. After the drop at 3.0 GPa to 4.0 GPa, the minimum thermal conductivity recovers at 5.0 GPa, reaching 3.23 Wm$^{-1}$K$^{-1}$. The observed drop in minimum thermal conductivity at 3.0 GPa and 4.0 GPa is consistent with the increase in anharmonic phonon interactions and phonon scattering due to the heightened Grüneisen parameter and thermal expansion coefficient, even though the phonon dynamics confirm that BeB$_2$C remains structurally stable throughout the pressure range. This highlights that stability does not preclude changes in thermal transport properties, as increased anharmonicity and local lattice adjustments can temporarily disrupt phonon flow, causing a decrease in thermal conductivity without leading to structural instability or phase transitions. Once the pressure surpasses 4.0 GPa, the lattice of the material stabilizes again with respect to phonon behavior, leading to a recovery in thermal conductivity at 5.0 GPa.

(d) *Heat capacity*

Utilizing the atom density per unit volume, $N$, the volumetric heat capacity at elevated temperatures is expressed as:

$$\rho C_P = c_p = 3Nk_B \tag{39}$$

where $k_B$ is Boltzmann's constant. The high value of heat capacity in a material indicates high value of thermal conductivity and low value of thermal diffusivity [67,68]. The heat capacity per unit volume of BeB$_2$C under different pressure was listed in Table 10.

(e) *Kleinman parameter* ($\zeta$)

The Kleinman parameter serves as an indicator of the phenomena pertaining to the elongation and deformation of atomic bonds. Specifically, $\zeta = 0$ denotes a condition in which the bending of bonds is minimized, whereas $\zeta = 1$ reflects a state in which the stretching of bonds is minimized. The Kleinman parameter delineates the relative geometrical arrangements of the cationic and anionic sublattices under volume-conserving strain distortions, wherein the positions are not constrained by symmetry considerations [69]. This observed stability implies that the atomic configuration of BeB$_2$C remains largely unaffected by substantial internal distortions when subjected to applied pressure [70]. The strain dependent Kleinman parameter was calculated using the following formula:

$$\zeta = \frac{C_{11} + 8C_{12}}{7C_{11} + 2C_{12}} \tag{40}$$

In Table 10 $\zeta$ increases from 0.53 at 0 GPa to 0.63 at 3 and 4 GPa, showing that bond bending becomes more significant under pressure, implying a change in the material's elastic behavior, where the structure responds more by changing bond angles rather than lengths, particularly around 3 GPa. However, at 5 GPa, it returns to 0.53, implying a return to the original balance of bond bending and stretching as pressure continues to rise.

(f) *Grüneisen parameter* ($\gamma$)

The Grüneisen parameter ($\gamma$) is a key characteristic in lattice dynamics, as it measures the anharmonicity of interatomic forces within a crystal structure. Anharmonicity refers to the deviations from the ideal harmonic approximation, where atomic vibrations are not perfectly symmetrical or linear. The Grüneisen parameter essentially provides insight into how the phonon frequencies of atoms within a solid change with volume or temperature. The parameter is calculated using the Poisson's ratio [71]:

$$\gamma = \frac{3}{2}\left(\frac{1+n}{2-3n}\right) \tag{41}$$

From Table 10 it can be seen that, $\gamma$ starts at 1.25 for 0 GPa, slightly increases to 1.29 up to 2 GPa, and peaks at 1.50 at 3 GPa. This suggests a maximum anharmonic behavior at 3 GPa, where lattice vibrations are more sensitive to volume changes. After 3 GPa, $\gamma$ drops back to 1.25 at 5 GPa, indicating reduced anharmonicity.

(g) *Thermal expansion coefficient (α)*

Another important thermophysical parameter, the thermal expansion coefficient (TEC), denoted by $\alpha$, measures how much a material expands or contracts with temperature change. It plays a crucial role in various physical properties and practical applications, especially in materials used for electronics, semiconductors, and crystal growth. The TEC is also intertwined with thermal and mechanical properties, making it a vital parameter for understanding and optimizing material behavior under varying thermal conditions, particularly in advanced technological applications. The TEC of a solid is inversely proportional to the modulus of rigidity and can be estimated using the following relation [72]:

$$\alpha = \frac{1.6 \times 10^{-3}}{G} \tag{42}$$

From Table 10, α starting at $1.13 \times 10^{-5}$ K$^{-1}$ at 0 GPa, remains fairly stable, dropping slightly at 2 and 5 GPa ($1.11 \times 10^{-5}$ K$^{-1}$), but increases at 3 and 4 GPa, peaking at $1.36 \times 10^{-5}$ K$^{-1}$. This suggests an increased sensitivity of the material to thermal expansion near 3-4 GPa, consistent with the trends in $\zeta$ and $\gamma$.

**Table 10:** Melting temperature ($T_m$ in K), minimum thermal conductivity ($k_{min}$ in Wm$^{-1}$K$^{-1}$), heat capacity per unit volume ($\rho C_\rho$ in JK$^{-1}$m$^{-3}$), Kleinman parameter ($\zeta$), Grüneisen parameter ($\gamma$), thermal expansion coefficient ($\alpha$ in K$^{-1}$) of BeB$_2$C under pressure ($P$ in GPa).

| $P$ | $k_{min}$ | $\rho C_\rho (\times 10^6)$ | $\zeta$ | $\gamma$ | $\alpha (\times 10^{-5})$ | Ref. |
|---|---|---|---|---|---|---|
| 0.0 | 3.18 | 5.64 | 0.53 | 1.25 | 1.13 |  |
| 1.0 | 3.19 | 5.67 | 0.54 | 1.29 | 1.13 |  |
| 2.0 | 3.22 | 5.70 | 0.54 | 1.29 | 1.11 | [This work] |
| 3.0 | 2.93 | 5.73 | 0.63 | 1.50 | 1.36 |  |
| 4.0 | 2.94 | 5.76 | 0.63 | 1.50 | 1.36 |  |
| 5.0 | 3.23 | 5.78 | 0.53 | 1.25 | 1.11 |  |

Table 9 and 10 demonstrate a coherent relationship between mechanical, thermal, and lattice dynamical properties of BeB$_2$C under pressure.

### 3.8 Anisotropies in sound velocity

Materials with high stiffness (large elastic constants) can transmit mechanical waves more quickly, while lower density reduces the inertia of the material, further allowing sound to travel faster The sound velocity in the [100] and [001] directions in the hexagonal structure can be calculated by the following equations [73]:

$$[100]v_l = \sqrt{C_{11} - C_{12}/2\rho} \; ; \quad [010]v_{t1} = \sqrt{C_{11}/\rho} \; ; \quad [001]v_{t2} = \sqrt{C_{44}/\rho}$$

$$[001]v_l = \sqrt{C_{33}/\rho} \; ; \qquad [100]v_{t1} = \sqrt{C_{44}/\rho} \; ; \qquad [010]v_{t2} = \sqrt{C_{44}/\rho}$$

where $\rho$ is the density of BeB$_2$C compound; $v_l$ is the longitudinal sound velocity; $v_{t1}$ and $v_{t2}$ refer to the first transverse mode and the second transverse mode, respectively. The calculated sound velocities in these directions for BeB$_2$C compounds are shown in Table 11.

**Table 11** Anisotropic sound velocities (in m/s) of BeB$_2$C system along different crystallographic directions (in m$^{-1}$).

| Propagation directions | Sound velocities |
|---|---|
| $[100]v_l$ | 1800.90 |
| $[010]v_{t1}$ | 3259.14 |
| $[001]v_{t2}$ | 2243.32 |
| $[001]v_l$ | 5158.04 |
| $[100]v_{t1}$ | 2243.32 |
| $[010]v_{t2}$ | 2243.32 |

The anisotropic sound velocities from Table 11 demonstrate the directional dependence of the elastic properties of BeB$_2$C, with the [001] direction being significantly stiffer than the [100] direction.

### 3.9 Optical properties

The response of a material to the incident electromagnetic radiation can be analyzed from the frequency dependent optical parameters. The study of the optical properties of solids reveals important facets of band structure, localized charged defects, lattice vibrations and impurity levels. In the recent years, with the increasing advancement and demand of the optoelectronic and photovoltaic devices, the study of different energy dependent optical properties of the materials has become a crucial part of fundamental materials science, technological innovation, with broad implications across multiple industries and scientific disciplines. In this section, we have investigated the optical properties such as absorption coefficient $\alpha(\omega)$, dielectric function $\varepsilon(\omega)$, photoconductivity $\sigma(\omega)$, refractive index n($\omega$), reflectivity R($\omega$) and loss function L($\omega$) in the photon energy range up to 30 eV with two different electric field polarization directions of [100] and [001]. A Gaussian smearing of 0.5 eV, a Drude energy 0.05 eV and an unscreened plasma energy of 5 eV were used to calculate the optical parameters as a function of incident photon energy. The complex dielectric function is given by:

$$\varepsilon(\omega) = \varepsilon_1(\omega) + i\varepsilon_2(\omega) \tag{43}$$

The imaginary part was obtained from Equation 2 and the real part was obtained from the Kramers-Kronig relationships. The imaginary part of the dielectric function describes the absorption of the incident electromagnetic radiation induced by interband electronic transitions and can reflect the energy band structure and other spectral information. Knowledge of the complex dielectric function allows the determination of different optical parameters using the following relations [74-76]:

$$n(\omega) = \frac{1}{\sqrt{2}} \left[ \sqrt{\varepsilon_1^2(\omega) + \varepsilon_2^2(\omega)} + \varepsilon_1(\omega) \right]^{1/2} \tag{44}$$

$$k(\omega) = \frac{1}{\sqrt{2}} \left[ \sqrt{\varepsilon_1^2(\omega) + \varepsilon_2^2(\omega)} - \varepsilon_1(\omega) \right]^{1/2} \tag{45}$$

$$R(\omega) = \left| \frac{\sqrt{\varepsilon(\omega)} - 1}{\sqrt{\varepsilon(\omega)} + 1} \right|^2 \tag{46}$$

$$\sigma(\omega) = \sigma_1(\omega) + i\sigma_2(\omega) = -i\frac{\omega}{4\pi}[\varepsilon_1^2(\omega) + \varepsilon_2^2(\omega) - 1] \tag{47}$$

$$\alpha(\omega) = \sqrt{2}\omega \left[ \sqrt{\varepsilon_1^2(\omega) + \varepsilon_2^2(\omega)} - \varepsilon_1(\omega) \right]^{1/2} \tag{48}$$

and,

$$L(\omega) = \frac{\varepsilon_2(\omega)}{\varepsilon_1^2(\omega) + \varepsilon_2^2(\varepsilon)} \tag{49}$$

The refractive index, denoted as $n(\omega)$, represents the ratio of the velocity of light in a vacuum to the velocity of light within a given material, thereby serving as an indicator of the extent to which the speed of electromagnetic radiation is reduced in a substance. The real part, $n(\omega)$, and the imaginary component, $k(\omega)$, of the refractive index for $BeB_2C$ at 0 GPa are depicted in Figure 9(a) and Figure 9(b), respectively. The imaginary part, $k(\omega)$, is commonly referred to as the extinction coefficient. The extinction coefficient $k(\omega)$ is intrinsically linked to the degree of light absorption by a material at a specified photon energy, demonstrating that elevated values of $k(\omega)$ are indicative of enhanced absorption [77]. Analysis of Figure 9(a) reveals that at lower photon energies, the refractive index tends to be relatively high, a phenomenon attributed to the more pronounced contribution of the material's electronic polarizability in decelerating the light. Around 5-10 eV, the difference is most prominent, with the [100] direction showing a higher refractive index. As the photon energy increases, it is customary for the refractive index to exhibit a downward trend. Furthermore, examination of Figure 9(b) illustrates that the extinction coefficient, $k(\omega)$, diminishes as the incident photon wavelength increases, a trend that correlates with the attenuation, or damping, of the oscillation amplitude of the incident electric field. With an increase in photon energy, both the refractive index and extinction coefficients exhibit oscillatory behavior. These properties make the material ideal for applications requiring directional control of light, such as in polarization-sensitive optical devices and photodetectors.

The reflectivity, denoted as $R(\omega)$, serves as an indicator of a material's capacity to reflect radiation, defined as the ratio of the intensity of reflected light to that of incident light [78]. The derived reflectivity of the $BeB_2C$ superconductor at 0 GPa is presented in Figure 9(c) for polarization directions [001] and [100]. The value of $R(\omega)$ initiates at approximately 98%, signifying that $BeB_2C$ reflects nearly all incident light, with minimal absorption or transmission occurring at lower photon energies. The graph illustrates a gradual decline in reflectivity of this compound as photon energy increases. From Figure 9(c), it is apparent that in the infrared and visible regions of the electromagnetic spectrum, reflectivity remains elevated and experiences a sharp decline within the ultraviolet region, ultimately approaching zero after 30 eV, accompanied by a distinct peak at approximately 3.50 eV along the [001] polarization direction. This observation provides further corroboration for the hypothesis that the pronounced reflectivity at lower photon energies suggests metallic characteristics, wherein the free electrons of the material effectively reflect light, while the material may exhibit increased transmissivity or absorptivity in regions characterized by lower reflectivity. This behavior may be associated with specific electronic transitions that facilitate light absorption. Consequently, $BeB_2C$ holds potential utility in applications where the regulation or maximization of light reflection is of paramount importance.

The optical conductivity of the $BeB_2C$, contingent upon photon energy at a pressure of 0 GPa, is illustrated in Figure 9(d) for the polarization directions of [001] and [100]. The figure indicates that the photoconductivity initiates from zero photon energy, signifying the ability of the material to conduct electricity even at minimal photon energy levels, subsequently increasing with elevated photon energy. This observation further corroborates the band structure analysis, which reveals the absence of a band gap in the material. Notably, in the visible and ultraviolet regions, peaks in conductivity suggest that $BeB_2C$ exhibits enhanced electrical conductivity, attributable to electronic transitions or an augmentation in carrier mobility. As demonstrated in Figure 9(d), the conductivity progressively diminishes within the ultraviolet region, ultimately approaching zero beyond 30 eV, a characteristic behavior commonly observed in metallic compounds.

The interaction of materials with incident electromagnetic radiation is quantitatively characterized by the frequency-dependent complex dielectric function $\varepsilon(\omega)$. The real part, $\varepsilon_1(\omega)$, and the imaginary part, $\varepsilon_2(\omega)$, of the dielectric function are shown in Figures 9(e) and 9(f), respectively. Analysis of Figure 9(e) reveals that $\varepsilon_1(\omega)$ begins from negative values of photon energy, exhibit anisotropy in two directions and indicative of the metallic characteristics of $BeB_2C$, which can be attributed to plasmonic phenomena. The real part shifts from negative to positive values, subsequently reverting to negative, and ultimately diminishing in the high-energy domain also reveals metallic nature of $BeB_2C$. The imaginary part of the dielectric function provides insights into the rate of absorption of electromagnetic waves, which is intrinsically linked to the material's capacity to absorb such waves [79]. From Figure 9(f), it is evident that $\varepsilon_2(\omega)$ exhibits substantial positive values at lower energy levels, signifying that $BeB_2C$ demonstrates high absorption for both polarization orientations, with peaks in this function correlating to specific photon energies where the material exhibits strong light absorption due to electronic transitions. Moreover, this reinforces the assertion that damping is predominantly characterized by $\varepsilon_2(\omega)$, with its influences manifesting in $\varepsilon_1(\omega)$ due to the interrelated dynamics of material interactions with electromagnetic radiation.

The absorption coefficient serves as a metric for quantifying the attenuation of electromagnetic radiation intensity per unit distance traversed within a given material, thereby influencing the interaction of the material with light and affecting its optical properties, thermal regulation, coloration, and applicability in various domains. Additionally, by analyzing the absorption coefficient as a function of photon energy, one can readily infer whether a material is a metal, semiconductor, or non-metal [80]. The absorption coefficient, $\alpha(\omega)$, of $BeB_2C$ at 0 GPa pressure along the polarization directions of [001] and [100] is depicted in Figure 9(g). Examination of Figure 9(g) reveals that the absorption coefficient initiates at approximately ~0.01 eV, with distinct peaks in the absorption coefficient at specific photon energies. The

α(*ω*) was quite high in the regions from about 10.67 eV to 17.02 eV peaking at ~15.48 eV belonging to the UV spectrum for [001] polarization direction.

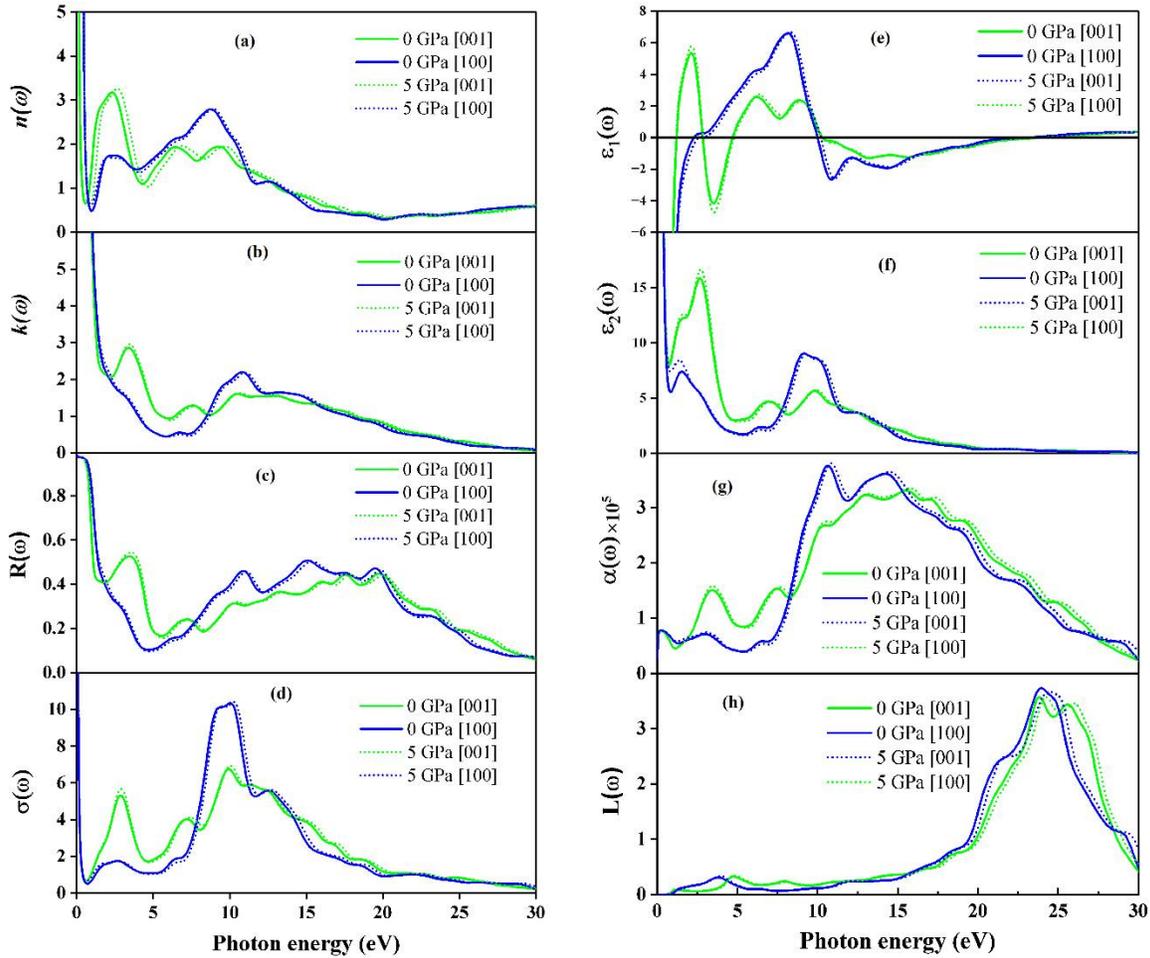

**Figure 9** The frequency dependent (a) real and (b) imaginary parts of the refractive index, (c) reflectivity, (d) optical conductivity, (e) imaginary and (f) imaginary parts of dielectric function, (g) absorption coefficient and (f) loss function of BeB$_2$C for the [100] and [001] electric field polarization directions at 0 GPa and 5 GPa.

The energy-loss function L(*ω*) represents a critical optical parameter, elucidating the manner in which plasmonic resonances and other collective excitations facilitate the dissipation of energy from electromagnetic waves as they propagate through a medium [81]. Figure 9(h) illustrates the frequency-dependent energy loss function of BeB$_2$C for the [001] and [100] polarization axes. The pronounced peaks within the loss function are indicative of plasma oscillations corresponding to specific photon energy, at which juncture the real component of the dielectric constant approaches zero. This peak is emblematic of plasmon resonance, characterized by the collective oscillation of the free electron gas present within the material. The energy level at which the loss function attains its peak is designated as the plasmon energy. Beyond this resonance energy, BeB₂C is anticipated to exhibit transparency to incident photons, thereby transitioning its response from metallic to dielectric-like behavior. The variations in peak positions and intensities serve to reflect the anisotropic nature of the material's electronic structure. It has been observed that the primary peak of L($\omega$) for the [001] and [100] orientations is situated at approximately 23.80 eV and 23.90 eV, respectively, under conditions of 0 GPa.

The plots reveal no significant change in optical properties when increasing pressure from 0 GPa to 5 GPa, implying that pressure does not induce phase transitions or substantial modifications in the material's electronic structure. So, we only provide the data for 0 GPa and 5 GPa. However, the optical characteristics do exhibit anisotropy concerning the polarization of the electromagnetic waves, as evidenced by the distinct behaviors in the [001] and [100] directions. This anisotropy suggests directional dependence in how the material interacts with light, which is typical of materials with complex electronic structures and/or crystal anisotropy. The results imply that despite pressure changes, the material remains structurally stable and does not undergo any electronic or phase transformations that could alter its optical response.

### 3.10 Phonon dynamics

By using the linear perturbative approach, we have calculated the phonon dispersion spectra (PDS) and phonon density of states (PHDOS) of BeB₂C compound along the high symmetry directions in the first Brillouin zone (BZ) at 0 GPa and 5 GPa pressures which have been shown in Figure 10.

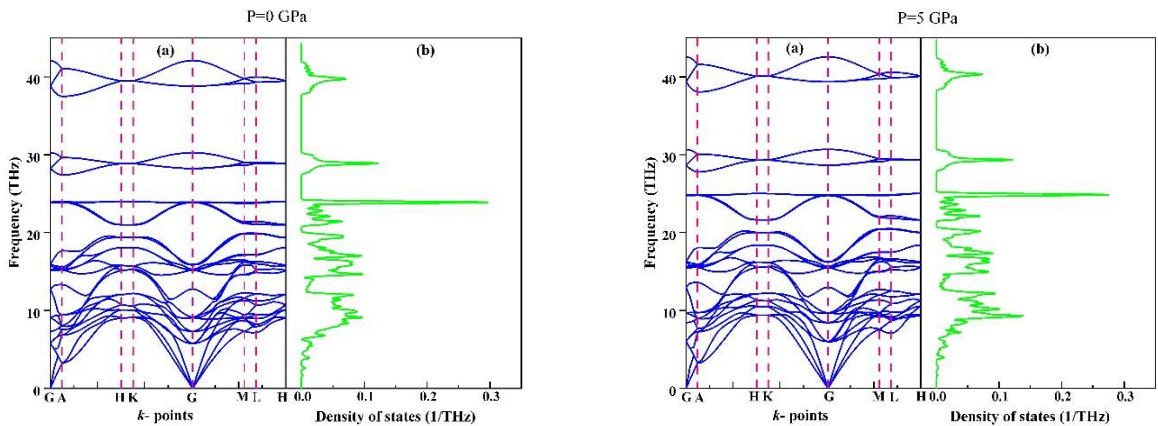

**Figure 10** Calculated (a) phonon dispersion spectra and (b) the phonon density of states for BeB₂C compound at 0 GPa and 5 GPa.

Pressure dependence of phonon dispersion relations can reveal important information about the bonding and structural features of a material and confirming the dynamical stability. The acoustic modes exhibit a phonon frequency of zero at the $\Gamma$-point, which serves as evidence of the dynamical stability of BeB₂C. From our calculations of phonon properties, we see that BeB₂C is dynamically stable when the pressure increases up to 5 GPa and there is no branch softening to imaginary frequencies. It is obvious that the contribution of the pressure is not significant. The acoustic phonons are created for in-phase movements of the atoms of the lattice about their equilibrium positions. On the other hand, optical phonons are out of phase movements of the atoms in the lattice and optical properties of crystals are mainly controlled by the optical branches. The prominent peaks in the PHDOS are observed around 24 THz frequency. It is seen that the high-frequency phonon modes (> 30THz) mainly result from the B atoms, whereas the low frequency phonon modes (1-20 THz) come from both Be and C atoms. Although Be is lighter than B, the nature of bonding in the lattice causes Be atoms to contribute more to lower-frequency modes, with the highest frequencies coming from the B atoms due to the strong B–B and B–C bonds in this compound. From Figure 10 it can be seen that, the flatness of the bands produces the peaks in the PHDOS and highly dispersive bands decrease the heights of peaks in the total PHDOS. These phonons are expected to play significant role in determining the optical properties of $BeB_2C$ [16,82]. From Figure 10 we can infer that, the $BeB_2C$ compound is dynamically stable under pressure and has promising thermal properties. Despite extensive studies on the structural and electronic properties of $BeB_2C$, no previous research has explored the influence of pressure on its phonon dispersion relations. This work addresses this gap, offering new insights into the dynamical stability and vibrational characteristics of the material at elevated pressures.

## 3.11 Superconductivity

The Bardeen-Cooper-Schrieffer (BCS) [83] theoretical framework is fundamentally based on electron-phonon interaction as the primary mechanism facilitating the formation of Cooper pairs. It is unequivocally acknowledged that the most prominent characteristic of $BeB_2C$ is its anticipated high superconducting critical temperature. $BeB_2C$ operates as an anisotropic two-gap electron–phonon superconductor, and its superconducting mechanism bears resemblance to that of $MgB_2$, which is a well-established two-gap electron-phonon superconductor exhibiting a high critical temperature of 39 K [84]. In this investigation, we have utilized the pseudopotential-based formulation proposed by McMillan, which is instrumental in calculating several crucial superconducting state parameters, including the electron-phonon coupling strength, $\lambda$, the Coulomb pseudopotential, $\mu^*$, the transition temperature, $T_c$, and the specific heat coefficient, employing pressure-dependent values of $\theta_D$ and $N(E_F)$. The superconducting transition temperature constitutes a pivotal superconducting parameter, which is theoretically derived via the McMillan formula [85] presented as follows:

$$T_c = \frac{\theta_D}{1.45} exp\left[\frac{-1.04(1+\lambda)}{\lambda - \mu^*(1+0.62\lambda)}\right] \qquad (50)$$

here, $\theta_D$ is the Debye temperature and $\mu^*$ refers to the repulsive screened Coulomb part (also known as the Coulomb pseudopotential) of the material, which were already listed in Table 7.

Herein, $N(E_F)$ denotes the value of the total density of states (TDOS) and Coulomb pseudopotential of the compound at the Fermi level has been previously addressed in the density of states (DOS) section. A positive electron-phonon coupling constant ($\lambda$) engenders an attractive interaction among electrons, which is critical for the formation of Cooper pairs in conventional superconductors.

Given that CASTEP is incapable of determining the Eliashberg spectral function, we have resorted to the electron-phonon coupling constants documented in prior research [85,86]. We used the value of $\lambda$ =

0.8036 by Zhang *et al* [5] just in order to approximately estimate the superconducting transition temperatures. For a fixed value of $\lambda$, $T_c$ is positively correlated with $\theta_D$. At 0 GPa, the calculated value $T_c \approx 26.6$ K which is very close to the value (30.2 K) found by Zhang *et al* [5]. As the $N(E_F)$ exhibits a slight variation (Table 6) with increasing pressure, this affects the value of $\mu^*$. From the McMillan equation, it is expected that for a fixed value of $\lambda$ of the BeB$_2$C compound, $T_c$ may increase with the applied pressure due to the increasing trend of $\theta_D$ for 2 GPa and 5 GPa. At 3 and 4 GPa $T_c$ might drop due to the decreasing trend of $\theta_D$ as shown in Table 9. This is because $\theta_D$ is linearly associated with $T_c$.

## 4. Conclusions

In conclusion, we have carried out a first-principles DFT exploration of the physical properties of BeB$_2$C. The material exhibits anisotropic compression, particularly along the *c*-axis, which makes it stiffer with increased pressure, while its six independent elastic constants confirm stability per Born-Huang criteria. Bulk and Young's moduli trends underscore its resistance to deformation, while ductility fluctuates with pressure, becoming brittle above 5 GPa. High hardness and machinability further validate its practical applicability. Phonon and elastic analyses confirm structural stability, while a metallic band structure and Fermi surface configuration suggest promising conductivity, though slightly diminished under pressure. It is noted that the B–B bonds manifest the highest degree of covalency and supports stability when juxtaposed with other bonding interactions, whereas the Be−B bond displays a partially ionic and comparatively diminished covalent nature within the BeB$_2$C compound. Thermophysical analysis reveals pressure impacts on Debye temperature and melting temperature showing potential for high-temperature applications. The Grüneisen parameter reveals that the peak anharmonicity and covalency manifest at pressures of 3 GPa and 4 GPa. Optical characteristics reveal strong anisotropic responses, with significant absorption in the UV region, marked by plasmon resonance around 23.8 eV. The compound reflects infrared and visible light very efficiently. It has potential to be used as a reflecting coating in optical devices. The computed superconducting transition temperature agrees well with previous finding [5]. We predict an insignificant variation of the superconducting transition temperature of BeB$_2$C within the pressure range considered in this study. Our findings underscore the versatility and potential of BeB$_2$C for high-performance applications. We hope this study will inspire the researchers to explore the physical properties of BeB$_2$C in greater depth in near future.

**Acknowledgements**
S.H.N. and R.S.I. acknowledge the research grant (1151/5/52/RU/Science-07/19-20) from the Faculty of Science, University of Rajshahi, Bangladesh, which partly supported this work. R.A. gratefully acknowledges the National Science and Technology (NST) Fellowship, awarded by the Ministry of Science and Technology, Bangladesh, for supporting his M.Sc. research. This work is dedicated to the cherished memory of the martyrs of the July-August 2024 movement in Bangladesh, whose sacrifices will forever inspire us.

**Data availability**
The data sets generated and/or analyzed in this study are available from the corresponding author on reasonable request.

**List of references**

[1] D. R. Glasson and J. A. Jones, Formation and reactivity of borides, carbides and silicides. I. Review and introduction, J. Appl. Chem. **19**, 125 (1969).

[2] P. D. Pancharatna, S. H. Dar, and M. M. Balakrishnarajan, Bonding in Boron Rich Borides, in Comprehensive Inorganic Chemistry III (Elsevier, 2023), pp. 26-50.

[3] S. H. Dar, P. D. Pancharatna, and M. M. Balakrishnarajan, Nature of Interactions between Boron Clusters: Extended Delocalization and Retention of Aromaticity Post-Oxidation, Inorg. Chem. **62**, 7566 (2023).

[4] P. Rogl and H. Bittermann, Ternary metal boron carbides, International Journal of Refractory Metals & Hard Materials **17** (1999) 27-32.

[5] D. Zhang, L. Feng, R. Wang, and Y. Shang, A New Superconductor of $BeB_2C$ Under Atmospheric Pressure, J Supercond Nov Magn **35**, 3135 (2022).

[6] H. M. Tütüncü, E. Karaca, and G. P. Srivastava, Electron-phonon interaction and superconductivity in the borocarbide superconductor, Philosophical Magazine **97**, 2669 (2017).

[7] J. Pan, B. Zhang, Y. Hou, T. Zhang, X. Deng, Y. Wang, N. Wang, and P. Sheng, Superconductivity in Boron-Doped Carbon Nanotube Networks.

[8] D. Li, C. Luo, and H. Wang, Studies of electronic structures and optical properties for rhombohedral $CeAlO_3$ and $PrAlO_3$ under pressure, Ferroelectrics **603**, 116 (2023).

[9] B. B. Karki, D. B. Ghosh, and S. K. Bajgain, Simulation of Silicate Melts Under Pressure, in Magmas Under Pressure (Elsevier, 2018), pp. 419-453.

[10] F. Parvin and S. H. Naqib, Pressure dependence of structural, elastic, electronic, thermodynamic, and optical properties of van der Waals-type $NaSn_2P_2$ pnictide superconductor: Insights from DFT study, Results in Physics **21**, 103848 (2021).

[11] R. A. Friesner, *Ab initio* quantum chemistry: Methodology and applications, Proc. Natl. Acad. Sci. U.S.A. **102**, 6648 (2005).

[12] G. Tse and D. Yu, The first principle study of electronic and optical properties in rhombohedral $BiAlO_3$, Mod. Phys. Lett. B **30**, 1650006 (2016).

[13] C. Li, B. Wang, R. Wang, H. Wang, and X. Lu, First-principles study of structural, elastic, electronic, and optical properties of hexagonal $BiAlO_3$, Physica B: Condensed Matter **403**, 539 (2008).

[14] R. Hill, Elastic properties of reinforced solids: Some theoretical principles, Journal of the Mechanics and Physics of Solids **11**, 357 (1963).

[15] Y.L. Wan, Q.D. Hao, C.E. Hu, Y. Cheng, and G.F. Ji, First-principles study of structural, electronic, elastic, vibrational and thermodynamics properties of $LaX_3$ ($X$ = Sn, Tl, Pb and Bi) compounds under pressure, Solid State Communications **336**, 114431 (2021).

[16] A. S. M. M. Reza and S. H. Naqib, *Ab-initio* investigation of the physical properties of BaAgAs Dirac semimetal and its possible thermo-mechanical and optoelectronic applications, Physica B: Condensed Matter **671**, 415425 (2023).

[17] S. Saha, T. P. Sinha, and A. Mookerjee, Electronic structure, chemical bonding, and optical properties of paraelectric $BaTiO_3$, Phys. Rev. B **62**, 8828 (2000).

[18] Md. Z. Rahaman and M. A. Islam, A Theoretical Investigation on the Physical Properties of $SrPd_2Sb_2$ Superconductor, J Supercond Nov Magn **34**, 1133 (2021).

[19] E. F. Talantsev, Advanced McMillan's equation and its application for the analysis of highly-compressed superconductors, Supercond. Sci. Technol. **33**, 094009 (2020).

[20] R. S. Mulliken, Electronic Population Analysis on LCAO-MO Molecular Wave Functions. I, The Journal of Chemical Physics **23**, 1833 (1955).

[21] F. Birch, Finite strain isotherm and velocities for single-crystal and polycrystalline NaCl at high pressures and 300°K, J. Geophys. Res. **83**, 1257 (1978).

[22] F. Mouhat and F.X. Coudert, Necessary and sufficient elastic stability conditions in various crystal systems, Phys. Rev. B **90**, 224104 (2014).

[23] S. T. Ahams, A. Shaari, R. Ahmed, N. F. A. Pattah, M. C. Idris, and B. U. Haq, *Ab initio* study of the structure, elastic, and electronic properties of $Ti_3(Al_{1-n}Si_n)C_2$ layered ternary compounds, Sci Rep **11**, 4980 (2021).

[24] M. E. Eberhart and T. E. Jones, Cauchy pressure and the generalized bonding model for nonmagnetic bcc transition metals, Phys. Rev. B **86**, 134106 (2012).

[25] S. Karkour, A. Bouhemadou, D. Allali, K. Haddadi, S. Bin-Omran, R. Khenata, Y. Al-Douri, A. Ferhat Hamida, A. Hadi, and A. F. Abd El-Rehim, Structural, elastic, electronic and optical properties of the newly synthesized selenides $Tl_2CdXSe_4$ ($X$ = Ge, Sn), Eur. Phys. J. B **95**, 38 (2022).

[26] D. Nguyen-Maxh, D. G. Pettifor, S. Znam, and V. Vitek, Negative Cauchy Pressure within the Tight-Binding Approximation, MRS Proc. **491**, 353 (1997).

[27] A. D. Sinnott, A. Kelly, C. Gabbett, M. Moebius, J. N. Coleman, and G. L. W. Cross, Pressure-dependent mechanical properties of thin films under uniaxial strain via the layer compression test, Journal of Materials Research **39**, 273 (2024).

[28] S. N. Tripathi, V. Srivastava, and S. P. Sanyal, Insight into Phase Transition, Electronic, Magnetic, Mechanical, and Thermodynamic Properties of TbTe: a DFT Investigation, J Supercond Nov Magn **31**, 3925 (2018).

[29] V. V. Brazhkin, High-pressure synthesized materials: treasures and hints, High Pressure Research **27**, 333 (2007).

[30] S. Al-Qaisi, M. S. Abu-Jafar, G. K. Gopir, R. Ahmed, S. Bin Omran, R. Jaradat, D. Dahliah, and R. Khenata, Structural, elastic, mechanical and thermodynamic properties of Terbium oxide: First-principles investigations, Results in Physics **7**, 709 (2017).

[31] Z. Hashin and S. Shtrikman, A variational approach to the theory of the elastic behaviour of polycrystals, Journal of the Mechanics and Physics of Solids **10**, 343 (1962).

[32] R. Hill, The Elastic Behaviour of a Crystalline Aggregate, Proc. Phys. Soc. A **65**, 349 (1952).

[33] S. F. Pugh, XCII. Relations between the elastic moduli and the plastic properties of polycrystalline pure metals, The London, Edinburgh, and Dublin Philosophical Magazine and Journal of Science **45**, 823 (1954).

[34] E. Maskar, A. F. Lamrani, M. Belaiche, H. Essaqote, A. Es-SMAIRI, T. V. Vu, and D. P. Rai, A DFT Study of Structural, Elastic, Thermodynamic, Magneto-optical, and Electrical Properties of Double-Perovskite $Bi_2CrMO_6$ ($M$ = Zn, Ni) Using GGA and TB-mBj Functionals, J Supercond Nov Magn **34**, 2105 (2021).

[35] I. N. Frantsevich, F. F. Voronov, and S. A. Bokuta, Elastic Constants and Elastic Moduli of Metals and Insulators Handbook, (1983).

[36] D. J. Quesnel, D. S. Rimai, and L. P. DeMejo, The Poisson ratio for an FCC Lennard-Jones solid, Solid State Communications **85**, 171 (1993).

[37] S. E. Birang O, H. S. Park, A. S. Smith, and P. Steinmann, Atomistic configurational forces in crystalline fracture, Forces in Mechanics **4**, 100044 (2021).

[38] F. Arab, F. A. Sahraoui, K. Haddadi, A. Bouhemadou, and L. Louail, Phase stability, mechanical and thermodynamic properties of orthorhombic and trigonal $MgSiN_2$: an *ab initio* study, Phase Transitions **89**, 480 (2016).

[39] E. Mazhnik and A. R. Oganov, A model of hardness and fracture toughness of solids, Journal of Applied Physics **126**, 125109 (2019).

[40] T. Guo, K. Liu, and R. Song, Crack propagation characteristics and fracture toughness analysis of rock-based layered material with pre-existing crack under semi-circular bending, Theoretical and Applied Fracture Mechanics **119**, 103295 (2022).

[41] P. Ravindran, L. Fast, P. A. Korzhavyi, B. Johansson, J. Wills, and O. Eriksson, Density functional theory for calculation of elastic properties of orthorhombic crystals: Application to $TiSi_2$, Journal of Applied Physics **84**, 4891 (1998).

[42] K. Liu, B. Dong, X. L. Zhou, S. M. Wang, Y. S. Zhao, and J. Chang, Structural, elastic, and thermodynamic properties of hexagonal molybdenum nitrides under high pressure from first principles, Journal of Alloys and Compounds **632**, 830 (2015).

[43] C. M. Kube, Elastic anisotropy of crystals, AIP Advances **6**, 095209 (2016).
[44] Y. H. Duan, Y. Sun, M. J. Peng, and S. G. Zhou, Anisotropic elastic properties of the Ca–Pb compounds, Journal of Alloys and Compounds **595**, 14 (2014).
[45] Md. A. Rahman, Md. Z. Rahaman, and Md. A. Rahman, The structural, elastic, electronic and optical properties of MgCu under pressure: A first-principles study, Int. J. Mod. Phys. B **30**, 1650199 (2016).
[46] O. Örnek, A. İyigör, A. S. Meriç, M. Çanlı, M. Özduran, and N. Arıkan, First-Principle Investigations of $(Ti_{1-x}V_x)_2FeGa$ Alloys. A Study on Structural, Magnetic, Electronic, and Elastic Properties, Russ. J. Phys. Chem. **95**, 2592 (2021).
[47] A. J. Crocker and L. M. Rogers, VALENCE BAND STRUCTURE OF PbTe, J. Phys. Colloques **29**, C4 (1968).
[48] Y. Bai, N. Srikanth, C. K. Chua, and K. Zhou, Density Functional Theory Study of $M_{n+1}AX_n$ Phases: A Review, Critical Reviews in Solid State and Materials Sciences **44**, 56 (2019).
[49] H. Yan, L. Chen, Z. Wei, M. Zhang, and Q. Wei, Superhard high-pressure structures of beryllium diborocarbides, Vacuum **180**, 109617 (2020).
[50] K. H. Bennemann, J. W. Garland, H. C. Wolfe, and D. H. Douglass, Theory for Superconductivity in D-Band Metals, in (Rochester, New York (USA), 1972), pp. 103-137.
[51] G. Jency and S. Sutha Kumari, *Ab Initio* Study on the Electronic Band Structure, Density of States, Structural Phase Transition and Superconductivity of Zirconium, Ujc **1**, 64 (2013).
[52] W. A. Dujana, A. Podder, O. Das, Md. Solayman, M. T. Nasir, M. A. Rashid, M. Saiduzzaman, and M. A. Hadi, Structural, electronic, mechanical, thermal, and optical properties of $UIr_3$ under pressure: A comprehensive DFT study, AIP Advances **11**, 105205 (2021).
[53] V. H. Tran and M. Sahakyan, Specific heat, Electrical resistivity and Electronic band structure properties of noncentrosymmetric $Th_7Fe_3$ superconductor, Sci Rep **7**, 15769 (2017).
[54] E. G. Michel, Fermi surface analysis using surface methods, J. Phys.: Condens. Matter **19**, 350301 (2007).
[55] M. I. Kholil and M. T. H. Bhuiyan, Elastic, Electronic, Optical, Thermodynamic, and Superconducting Properties of $CaMSi_3$ (*M* = Ir, Pt) and La*M*$Si_3$ (*M* = Ir, Rh) Superconductors: Insights from DFT-Based Computer Simulation, J Supercond Nov Magn **34**, 1775 (2021).
[56] F. L. Hirshfeld, Bonded-atom fragments for describing molecular charge densities, Theoret. Chim. Acta **44**, 129 (1977).
[57] Z. Demircioğlu, Ç. A. Kaştaş, and O. Büyükgüngör, Theoretical analysis (NBO, NPA, Mulliken Population Method) and molecular orbital studies (hardness, chemical potential, electrophilicity and Fukui function analysis) of (E)-2-((4-hydroxy-2-methylphenylimino)methyl)-3-methoxyphenol, Journal of Molecular Structure **1091**, 183 (2015).
[58] F. Gao, Theoretical model of intrinsic hardness, Phys. Rev. B **73**, 132104 (2006).
[59] W. Liu, Y. P. Zhou, and X. L. Feng, Hardness Prediction and First Principle Study of *Re*-123(*Re* = Y, Eu, Pr, Gd) Superconductors, Bulletin of the Korean Chemical Society **30**, 3016 (2009).
[60] F. M. Gao and L. H. Gao, Microscopic models of hardness, J. Superhard Mater. **32**, 148 (2010).
[61] Z.Y. Jiao, S.H. Ma, X.Z. Zhang, and X.F. Huang, Pressure-induced effects on elastic and mechanical properties of TiC and TiN: A DFT study, EPL **101**, 46002 (2013).
[62] O. L. Anderson, A simplified method for calculating the debye temperature from elastic constants, Journal of Physics and Chemistry of Solids **24**, 909 (1963).
[63] S. Daoud, N. Bioud, and N. Lebgaa, Elastic and piezoelectric properties, sound velocity and Debye temperature of (B3) boron-bismuth compound under pressure, Pramana - J Phys **81**, 885 (2013).
[64] M. E. Fine, L. D. Brown, and H. L. Marcus, Elastic constants versus melting temperature in metals, Scripta Metallurgica **18**, 951 (1984).
[65] N. B. Duc, H. K. Hieu, P. T. M. Hanh, T. T. Hai, N. V. Tuyen, and T. T. Ha, Investigation of melting point, Debye frequency and temperature of iron at high pressure, Eur. Phys. J. B **93**, 115 (2020).
[66] D. R. Clarke, Materials selection guidelines for low thermal conductivity thermal barrier coatings, Surface and Coatings Technology **163-164**, 67 (2003).

[67] I. Hatta, Heat capacity per unit volume, Thermochimica Acta **446**, 176 (2006).
[68] Nanomaterials, Nanotechnologies and Design (Elsevier, 2009).
[69] W. A. Harrison, Electronic Structure and the Properties of Solids: The Physics of the Chemical Bond (Courier Corporation, 2012).
[70] L. Kleinman, Deformation Potentials in Silicon. I. Uniaxial Strain, Phys. Rev. **128**, 2614 (1962).
[71] D. S. Sanditov and M. V. Darmaev, Effective Modulus of Elasticity and Grüneisen Parameter of Chalcogenide Glasses in the As–Tl–S System, Inorg Mater **55**, 617 (2019).
[72] M. F. Ashby and D. Cebon, Materials selection in mechanical design, J. Phys. IV France **03**, C7 (1993).
[73] W. Bao, D. Liu, and Y. Duan, A first-principles prediction of anisotropic elasticity and thermal properties of potential superhard $WB_3$, Ceramics International **44**, 14053 (2018).
[74] Y. Nassah, A. Benmakhlouf, L. Hadjeris, T. Helaimia, R. Khenata, A. Bouhemadou, S. Bin Omran, R. Sharma, S. Goumri Said, and V. Srivastava, Electronic band structure, mechanical and optical characteristics of new lead-free halide perovskites for solar cell applications based on DFT computation, Bull Mater Sci **46**, 55 (2023).
[75] D. Li, C. Luo, and H. Wang, Studies of electronic structures and optical properties for rhombohedral $CeAlO_3$ and $PrAlO_3$ under pressure, Ferroelectrics **603**, 116 (2023).
[76] A. Tasnim, A. Afzal, R. S. Islam, and S. H. Naqib, Pressure-dependent semiconductor-metal transition and elastic, electronic, optical, and thermophysical properties of SnS binary chalcogenide, Results in Physics **45**, 106236 (2023)
[77] B. Saleh and M. Teich, Fundamentals of Photonics, 3rd Edition (2019).
[78] T. E. Zavecz, M. A. Saifi, and M. Notis, Metal reflectivity under high-intensity optical radiation, Applied Physics Letters **26**, 165 (1975).
[79] T. Qiang, The imaginary part of dielectric function and the absorption coefficient, College Physics (2009).
[80] R. O. Akande and E. O. Oyewande, Photon absorption potential coefficient as a tool for materials engineering, Int Nano Lett **6**, 243 (2016).
[81] X. Zhang, H. Xiang, M. Zhang, and G. Lu, Plasmonic resonances of nanoparticles from large-scale quantum mechanical simulations, Int. J. Mod. Phys. B **31**, 1740003 (2017).
[82] Z. W. Niu, Y. Cheng, H.Y. Zhang, and G.F. Ji, First-Principles Investigations on Structural, Phonon, and Thermodynamic Properties of Cubic $CeO_2$, Int J Thermophys **35**, 1601 (2014).
[83] J. Bardeen, L. N. Cooper, and J. R. Schrieffer, Theory of Superconductivity, Phys. Rev. **108**, 1175 (1957).
[84] J. Nagamatsu, N. Nakagawa, T. Muranaka, Y. Zenitani, and J. Akimitsu, Superconductivity at 39 K in magnesium diboride, Nature **410**, 63 (2001).
[85] W. L. McMillan, Transition Temperature of Strong-Coupled Superconductors, Phys. Rev. **167**, 331 (1968).
[86] F. Giustino, Electron-phonon interactions from first principles, Rev. Mod. Phys. **89**, 015003 (2017).

**Author Contributions**

**Ruman Ali**: Methodology, Software, Writing- Original draft. **Md. Enamul Haque**: Methodology, Software. **Jahid Hassan**: Methodology, Software. **M. A. Masum**: Methodology, Software. **R. S. Islam**: Supervision, Writing-Reviewing and Editing. **S. H. Naqib**: Conceptualization, Supervision, Formal Analysis, Writing- Reviewing and Editing.

**Competing Interests**

The authors declare no competing interests.